\documentclass[a4paper,aps,pra,showpacs, twocolumn, nosuperscriptaddress]{revtex4-2} 

\usepackage[utf8]{inputenc}  
\usepackage[T1]{fontenc}     
\usepackage[american]{babel}  
\usepackage[colorlinks,citecolor=blue,linkcolor=magenta,urlcolor=blue]{hyperref}  
\usepackage{graphicx} 
\usepackage{amsmath,amssymb,amsthm,bm,mathtools,amsfonts,mathrsfs,bbm,dsfont} 

\newcommand{\bra}[1]{\langle #1|}
\newcommand{\ket}[1]{|#1\rangle}
\newcommand{\braket}[2]{\langle #1|#2\rangle}

\newcommand{\be}{\begin{equation}}
\newcommand{\ee}{\end{equation}}
\newcommand{\bea}{\begin{eqnarray}}
\newcommand{\eea}{\end{eqnarray}}

\newcommand{\va}[1]{\ensuremath{(\Delta#1)^2}}

\newcommand{\eins}{\mathbbm{1}}

\newcommand{\kommentar}[1]{}
\newcommand{\tr}{{\mathrm{tr}}}

\renewcommand{\vr}{\ensuremath{\varrho}}

\renewcommand{\vec}[1]{\ensuremath{\boldsymbol{#1}}}

\usepackage{graphicx, xcolor}
\usepackage{braket}

\begin{document}
\title{Work fluctuations and entanglement in quantum batteries
}
\author{Satoya Imai}
\author{Otfried G\"uhne}
\author{Stefan Nimmrichter}

\affiliation{Naturwissenschaftlich-Technische Fakult\"at, Universit\"at Siegen, Walter-Flex-Str.~3, D-57068 Siegen, Germany}

\date{\today}

\begin{abstract}
We consider quantum batteries given by composite interacting quantum systems in terms of the thermodynamic work cost of local random unitary processes.
We characterize quantum correlations by monitoring the average energy change and its fluctuations in the high-dimensional bipartite systems.
We derive a hierarchy of bounds on high-dimensional entanglement (the so-called Schmidt number) from the  work fluctuations and thereby show that larger work fluctuations can verify the presence of stronger entanglement in the system.
Finally, we develop two-point measurement protocols with noisy detectors that can estimate work fluctuations, showing that the dimensionality of entanglement can be probed in this manner.
\end{abstract}

\maketitle

\section{Introduction}
At the heart of quantum thermodynamics \cite{goold2016role} lies the 
fundamental question about the emergence of thermodynamic properties 
in small quantum systems. Quantum thermodynamics has not only established 
a common playground for statistical mechanics and quantum 
information theorists, it is now driving experimental efforts to seek and 
exploit genuine quantum signatures in thermodynamic processes. In particular, 
quantum correlations have been investigated in terms of their fundamental 
energetic footprint \cite{mukherjee2016presence, alimuddin2019bound} and 
work cost \cite{bruschi2015thermodynamics, perarnau2015extractable, huber2015thermodynamic, friis2016energetics, brunelli2017detecting,mckay2018fluctuations}, and as a resource
in quantum thermal machines \cite{brunner2014entanglement, brask2015autonomous}. Research on quantum batteries \cite{campaioli2018quantum} highlights the role of correlations for work extraction \cite{alicki2013entanglement, hovhannisyan2013entanglement, salvia2022optimal} and storage
\cite{binder2015quantacell, campaioli2017enhancing, friis2018precision, le2018spin, Ferraro2018highpower, Andolina2019extractablework, julia2020bounds, Quach2020usingdark, Gyhm2022quantumcharging} in composite quantum systems. 
{Experimental investigations of quantum batteries are already underway \cite{quach2022superabsorption,Hu2021Optimal}.}

In practice, if work is consumed or generated on the quantum scale, strong 
fluctuations are often inevitable. Whether they are caused by a lack of 
experimental control, environmental decoherence, or other unknown sources 
of noise, the fluctuations are not only detrimental to the performance of 
thermodynamic tasks, but their precise statistics are often inaccessible.
It is a common approach in quantum information theory to circumvent this 
problem by considering---or even deliberately applying---uniformly random 
unitary operations on the quantum system \cite{van2012measuring, tran2015quantum, tran2016correlations, elben2018renyi,elben2019statistical, brydges2019probing, elben2020cross, ketterer2019characterizing, ketterer2020entanglement, imai2021bound, ketterer2020certifying}. This operational 
``worst-case'' procedure will override other noise effects by rotating around an 
arbitrary Hilbert 
space direction, which results in a maximally mixed system state on 
average. Nevertheless, measurement data from a large sample of 
random unitaries can reveal genuine quantum features of the system 
state.

In the context of quantum thermodynamics, random unitaries and random Hamiltonians that generate them have been used to characterize the work distribution in chaotic quantum systems \cite{garcia2017quantum,lobejko2017work,chenu2018quantum,chenu2019work,salvia2021distribution}. 
Other studies analyzed the thermodynamics of quantum batteries under random 
unitary rotations \cite{caravelli2020random, oliviero2021random}, random 
repeated collisions \cite{gennaro2009relaxation,de2020quantum, Shaghaghi2021extracting}, or random interaction Hamiltonians \cite{rossini2020quantum,rosa2020ultra,jia2020spectral}.

\begin{figure}[t]
    \centering
    \includegraphics[width=1\columnwidth]{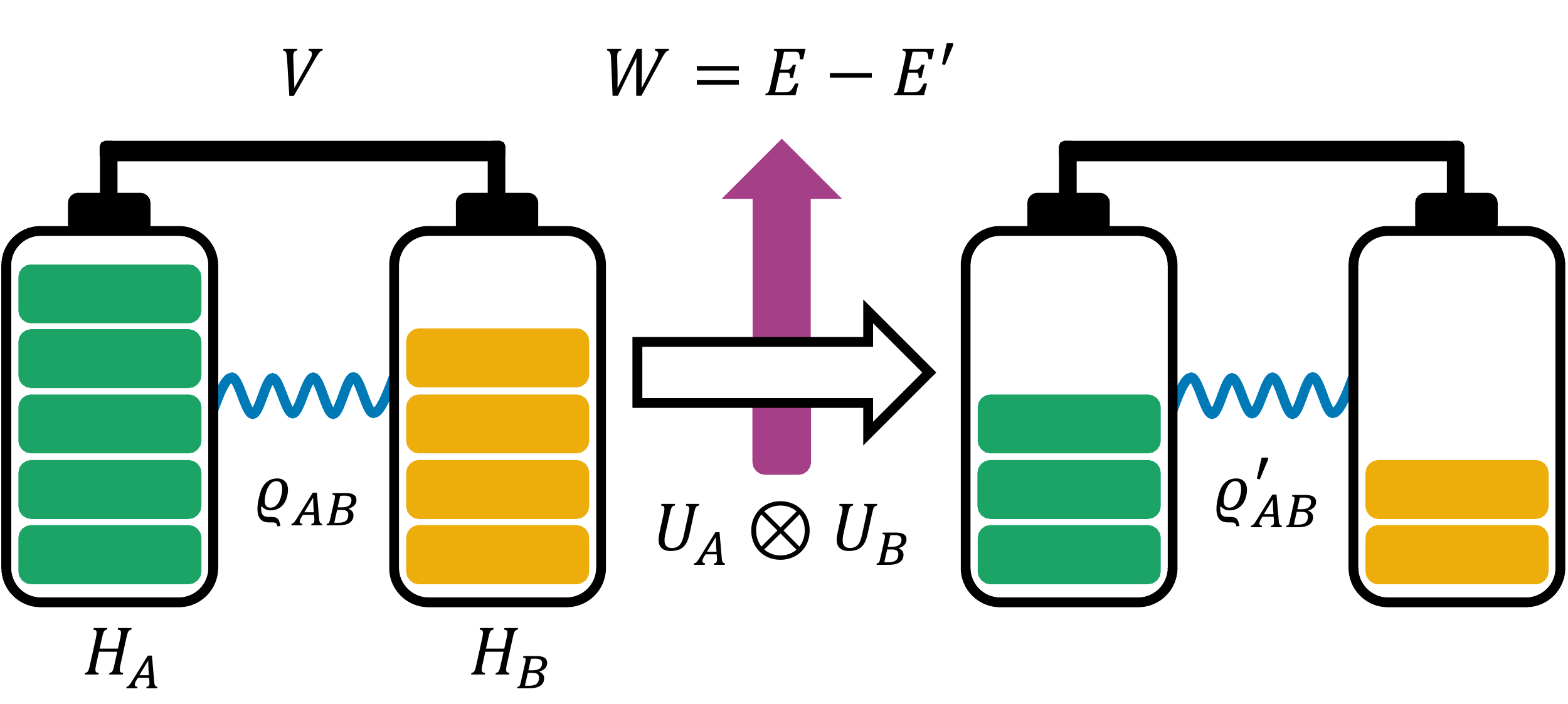}
    \caption{
    Sketch of the interacting quantum battery as a composite working medium 
    that can be entangled in a $d \times d$ system.
    The quantum battery is described by a state $\vr_{AB}$ and a Hamiltonian 
    $H_{AB} = H_A + H_B + g V$ with coupling strength $g$. It is transformed 
    by a local random unitary operation $U_A\otimes U_B$: $\vr_{AB} \to \vr_{AB}^\prime = (U_A \otimes U_B) \vr_{AB} (U_A^\dagger \otimes U_B^\dagger)$.
    Then the average extractable work in this process $W(U_A,U_B)=E-E^\prime$ becomes 
    random. The essential thermodynamic quantity to characterize high-dimensional entanglement in this paper is the work variance $\va{\overline{W}}$ over the random unitaries.
    }
    \label{sketch}
\end{figure}

Here, we show that one can detect bipartite entanglement in a composite interacting working medium through work fluctuations under \textit{local} random unitaries.
We derive a hierarchy of bounds on high-dimensional entanglement in terms of the so-called Schmidt number, and we show that stronger work fluctuations can verify the presence of stronger entanglement.
Furthermore, we develop noisy two-point energy measurement protocols based on inefficient detectors that can estimate work fluctuations and thereby probe the Schmidt number.

\section{Quantum battery}
Consider an interacting bipartite quantum system with dimension $d \times d$ and 
Hamiltonian $H_{AB} = H_A \otimes \eins_{B} + \eins_{A} \otimes H_B + g V$, 
prepared in a (possibly entangled) quantum state $\vr_{AB}$. Its energy content 
$E= \tr [\vr_{AB} H_{AB}]$ has contributions from the local Hamiltonians $H_A,H_B$ 
and from the interaction term $V$ at coupling strength $g$. The system shall act 
as a quantum battery that receives or delivers energy through a local {(non-entangling)} unitary control operation,
which we describe by $\vr_{AB}^\prime = (U_A \otimes U_B) \vr_{AB} (U_A^\dagger \otimes U_B^\dagger)$, see also Fig.~\ref{sketch}.
{We assume a pulsed (or cyclic) operation that leaves the system Hamiltonian unchanged, i.e., $H_{AB}^\prime=H_{AB}$.}

The associated locally extracted work is quantified by the energy difference
\begin{align} \label{extractablework}
     W(U_A,U_B) =E - E^{\prime} = \tr[ (\vr_{AB}-\vr_{AB}^\prime)H_{AB} ].
\end{align}
Most studies on quantum battery (dis-)charging focus on the maximum amount of 
the extractable work, called ergotropy \cite{allahverdyan2004maximal}, which has recently been linked to quantum correlations \cite{alimuddin2019bound, Francica2017daemonic, Bernards2019daemonic, Francica2022quantum}. In this 
paper, we will not be concerned with the maximization, but rather with the work statistics over a sample of uniformly random local operations and relate 
it to the entanglement between the parts of the battery. We consider 
the average work and its variance over a sample of unitaries $U_A, U_B$ 
drawn from the unitary groups $\mathcal{U}(d)$:
\begin{align}
\overline{W} &= \int dU_A \int dU_B  \, W(U_A,U_B), \label{eq:workMean} \\
\va{\overline{W}}&= \overline{W^2} - \overline{W}^2, \label{eq:workVariance}
\end{align}
where the integrals are taken over the Haar measure, {see Appendix \ref{app:A} for details, including a commented list of useful identities and known results.}
We immediately find that $\overline{W}=E-{\tr[H_{AB}]}/{d^2}$, since the averaged 
final battery state is always maximally mixed.
On the other hand, we will see that the variance $\va{\overline{W}}$ of work 
fluctuations can reveal initial quantum correlations in the battery.

\section{Work fluctuations}
By virtue of the Schur-Weyl duality \cite{goodman2000representations, roberts2017chaos, kliesch2021theory}, we can carry out the unitary integrals 
in Eq.~\eqref{eq:workVariance} and link the work fluctuations to the generalized 
Bloch decomposition of $\vr_{AB}$ and $H_{AB}$.
Recall that any $d \times d$ state $\vr_{AB}$ can be written as
\begin{align} \nonumber
    \vr_{AB} &= \frac{1}{d^2} \Big(
    \eins_{AB}
    + \sum_{i=1}^{d^2-1}r_i^{A} \lambda_i \otimes \eins_{B}\\
    &\quad
    + \sum_{i=1}^{d^2-1}r_i^{B} \eins_{A} \otimes \lambda_i
    + \sum_{i,j=1}^{d^2-1} t_{ij} \lambda_i \otimes \lambda_j
    \Big),
    \label{blochdecompos}
\end{align}
with $\lambda_{0}=\eins_{d}$ and $\lambda_{i}$ the so-called Gell-Mann matrices for $i=1,\ldots, d^2-1$
\cite{kimura2003bloch, bertlmann2008bloch, Siewert2022orthogonal}.
These matrices generalize the Pauli matrices to SU$(d)$, satisfying
$\lambda_i^\dagger = \lambda_i$, $\tr[\lambda_i] = 0$, and $\tr[\lambda_i \lambda_j] 
= d \delta_{ij}$. The coefficient vectors $\vec{r}^A$ and $\vec{r}^B$ characterize 
the two reduced battery states, while the matrix $(t_{ij})$ represents all 
correlations.
Similarly, we can expand the terms of the Hamiltonian as
\begin{align} \label{eq:HamExpansion}
    H_{X}    = \sum_{i=0}^{d^2-1} h_i^{X} \lambda_i,\,\,\,\,
    V   = \sum_{i,j=1}^{d^2-1} v_{ij} \lambda_i \otimes \lambda_j.
\end{align}
{This leads to an explicit form for the work fluctuations:}

{\bf Observation 1.}
\textit{
The work variance over local random unitary operations in a $d \times d$ quantum battery described by $\vr_{AB}$ and $H_{AB}$ can be written in terms of the Bloch representation as 
\begin{align}\label{eq:varW_ideal}
   \va{\overline{W}} = \frac{1}{d^2-1}\left(
   r_A^2 h_A^2 + r_B^2 h_B^2+ \frac{t^2 g^2v^2}{d^2-1}
   \right),
\end{align}
where
$r_X^2 = |\vec{r}^{X}|^2$,
$t^2 = \sum_{i,j} t_{ij}^2$,
$h_{X}^2 = |\vec{h}^{X}|^2$, and
$v^2   = \sum_{i,j} v_{ij}^2$,
for $X=A,B$.
}

\begin{proof}
First, we can immediately find
\begin{align}
    \va{\overline{W}} = \overline{(E^\prime)^2} - \overline{E^\prime}^2.
\end{align}
The first term on this right-hand side can be written as
\begin{align}
    \overline{(E^\prime)^2} \nonumber
    &= \int dU_A \int dU_B\,
    \left\{\tr[\vr_{AB}^\prime H_{AB}]\right\}^2\\ \nonumber
    &= \int dU_A \int dU_B\,\,
    \tr\left[\vr_{AB}^{\prime^{\otimes 2}} H_{AB}^{\otimes 2}\right]\\ \nonumber
    &= 
    \tr\left[
    \left(\int dU_A \int dU_B\, \vr_{AB}^{\prime^{\otimes 2}}\right)
    H_{AB}^{\otimes 2}\right]\\
    &= 
    \tr\left[
    \Phi(\vr_{AB}) H_{AB}^{\otimes 2}\right],
\end{align}
where the map $\Phi(\vr_{AB})$ is given in Appendix \ref{app:A}, see Eq.~(\ref{PhivrABmap}).
Using the expansion of the Hamiltonian terms in Eq.~\eqref{eq:HamExpansion}, with help of the properties of Gell-Mann matrices, a long but straightforward calculation leads to the expression \eqref{eq:varW_ideal}.
\end{proof}

Similar quantities have appeared in the notion of sector lengths in quantum
information theory \cite{aschauer2003local, de2011multipartite, klockl2015characterizing, wyderka2020characterizing, eltschka2020maximum}.
Here, the bipartite correlations of the battery state $\vr_{AB}$ contribute 
to $\va{\overline{W}}$ via the term $t^2$, provided there is a finite 
coupling $g\neq 0$ between the battery parts. Next, we will characterize 
the entanglement in $\vr_{AB}$ based on Eq.~\eqref{eq:varW_ideal}.

\section{Schmidt number detection}
A typical way to describe high-dimensional entanglement in a pure bipartite 
state $\ket{\psi}$ is to consider its Schmidt decomposition \cite{nielsen2002quantum},
$\ket{\psi}=\sum_{i=1}^{r} \sqrt{\lambda_i} \ket{e_i}\otimes \ket{f_i}$,
with $\braket{e_i|e_j}=\braket{f_i|f_j}=\delta_{ij}$ and $\sum_{i=1}^r \lambda_i=1$.
The number $r = r(\psi)$ is equal to the rank of $\tr_{A}[\ket{\psi}\!\bra{\psi}]$, and also called the Schmidt rank, and the state $\ket{\psi}$ is entangled 
iff $r(\psi)>1$. A high Schmidt rank certifies high-dimensional entanglement 
of the state, which may imply usefulness for certain information processing tasks \cite{cerf2002security, barrett2006maximally, buscemi2011entanglement,huber2013weak}.

The generalization of the Schmidt rank to mixed states $\vr_{AB}$ is known 
as the Schmidt number \cite{terhal2000schmidt}:
\begin{align}
    \mathrm{SN}(\vr_{AB}) =
    {\inf_{\mathcal{D}(\vr_{AB})}}
    \max_{\{{\psi_i}\}}\, r({\psi_i}),
\end{align}
where
$\mathcal{D}(\vr_{AB})
= \{p_i, \psi_i: \vr_{AB} = \sum_i p_i \ket{\psi_i} \bra{\psi_i} \}$ is the 
set of all ensemble realizations of $\vr_{AB}$. The sets $\mathcal{S}_k$ of 
all bipartite states with $\mathrm{SN}=k$ form a hierarchy of convex and 
compact subsets in state space, $\mathcal{S}_k \subset \mathcal{S}_{k+1}$, 
where $\mathcal{S}_1$ is the set of separable states. A higher Schmidt number 
thus indicates stronger entanglement, augmenting the separability problem \cite{guhne2009entanglement, friis2019entanglement}. Several methods to witness the Schmidt number are already known \cite{sanpera2001schmidt, sperling2011determination, sperling2011schmidt, huber2018high, Liu2022bounding}. We now formulate a criterion based 
on work fluctuations, which elucidates the role of entanglement in work 
exchange processes:

{\bf Observation 2.}
\textit{
Any $d \times d$ composite quantum battery described by $\vr_{AB}$ and $H_{AB}$ with $\mathrm{SN}(\vr_{AB})= k$ obeys
\begin{align}\label{SNdetect_from_WFs}
\va{\overline{W}} \leq
\frac{1}{d^2-1}\left(
   r_A^2 h_A^2 + r_B^2 h_B^2+ \frac{g^2v^2s_k}{d^2-1}
   \right),
\end{align}
with the function
$s_k=s\left(k,d,r_A^2, r_B^2\right)= kd-1 + \frac{kd-2}{2} \left(r_A^2+r_B^2\right)
- \frac{kd}{2}\left|r_A^2-r_B^2\right|$.}

\begin{proof}
Let us begin by considering a map given by
\begin{align}
    M_k (X) = \tr[X]\eins - \frac{X}{k},
\end{align}
for an operator $X \in \mathcal{H}^d$ and an integer $k$.
Ref.~\cite{terhal2000schmidt} showed that,
if a two-qudit state $\vr_{AB}$ has Schmidt number $\mathrm{SN}(\vr_{AB}) = k$, then
$(M_k \otimes \eins_B)(\vr_{AB})$ is positive, 
\begin{align}
    (M_k \otimes \eins_B)(\vr_{AB})
    = \vr_{A} \otimes \eins_B -\frac{1}{k}\vr_{AB} \geq 0,
\end{align}
where $\vr_{A}=\tr_B[\vr_{AB}]$.
Noting that $\tr[\vr_{AB} O] \geq 0$ for any positive operator $O$, and
taking $O= (M_k \otimes \eins_B)(\vr_{AB})$, we have
\begin{align}
    \tr[\vr_{AB}^2] \leq k \,  \tr[\vr_{A}^2].
\end{align}
Similarly, we can show that $\tr[\vr_{AB}^2] \leq k \,  \tr[\vr_{B}^2]$.
In summary, any $d \times d$ quantum state $\vr_{AB}$ with Schmidt number $k$ obeys
\begin{align}
    \tr[\vr_{AB}^2]
    \leq k 
    \min\left\{\tr[\vr_{A}^2], \, \tr[\vr_{B}^2]\right\}.
\end{align}
For $k=1$, this inequality becomes equivalent to the well-known entropic separability criterion \cite{horodecki1996information, elben2018renyi}.

Here we note that
\begin{align}
\label{PurityandSLs}
    \tr[\vr_{AB}^2] = \frac{1}{d^2}\left(
    1+r_A^2+r_B^2+t^2
    \right).
\end{align}
Using Eq.~(\ref{PurityandSLs}) and $\min(a,b)=(a+b-|a-b|)/2$, we can rewrite the above condition as
\begin{align} \label{snconditionT2}
t^2 \leq
kd-1 + \frac{kd-2}{2} \left(r_A^2+r_B^2\right)
- \frac{kd}{2}\left|r_A^2-r_B^2\right|.
\end{align}
In the above Observation, the right-hand side of \eqref{snconditionT2} is subsumed as $s_k \equiv s\left(k,d,r_A^2,r_B^2\right)$.
A violation of this inequality implies that the state has a Schmidt number of at least $(k+1)$.
Observation 2 follows by applying the inequality to the work fluctuations $\va{\overline{W}}$ in Eq.~\eqref{eq:varW_ideal}.
We remark that a similar proof technique was employed in Ref.~\cite{imai2021bound}.
\end{proof}

A violation of Eq.~\eqref{SNdetect_from_WFs} implies that the battery state $\vr_{AB}$ 
has a Schmidt number of at least $(k+1)$. Hence, observing stronger work 
fluctuations from local random unitaries on a composite quantum battery 
allows us to detect high-dimensional entanglement.

{Note that the converse argument can be also true in the case of pure states.
To see this, we begin by noting that the purity constraint $\tr[\vr_{AB}^2]=1$ is equivalent to
$
    r_A^2+r_B^2 =d^2-1 -t^2.
$
For the sake of simplicity, assuming
$h_A^2 = h_B^2 = h^2$,
we can then express $\va{\overline{W}}$ as
\begin{align}
   \va{\overline{W}} =
   h^2 +  \frac{Gt^2}{d^2-1},
\end{align}
where
$
    G = ({g^2v^2})/({d^2-1})-h^2.
$
Also, we can rewrite the Schmidt number criterion  as
$
    t^2 \leq d^2 +1 - \frac{2d}{k}.
$
If the interaction is sufficiently strong, that is, $G>0$, then we get an upper bound on $\va{\overline{W}}$ from the Schmidt number criterion and arrive at the same conclusion as Observation 2.
On the other hand, if the interaction is weak, $G<0$, then a \textit{lower} bound on $\va{\overline{W}}$ is obtained, and hence \textit{weaker} work fluctuations would certify higher entanglement.

We remark that our approach to detect high-dimensional entanglement by observing random fluctuations can be applied not only to energy, but also to other observables measuring bipartite correlations.
}

\section{Example}
We shall test our criterion with the family of states
\begin{align}\label{eq:thermalWstates}
    \vr_{\alpha} = \alpha \ket{\phi}\!\bra{\phi}
    +(1-\alpha)\tau_A \otimes \tau_B.
\end{align}
They are mixtures between the product of local Gibbs states at temperature $T$, $\tau_X=\exp(-H_X/T)/Z_X$, and the pure entangled state $\ket{\phi}$ that is locally indistinguishable from the Gibbs states, $\tr_A(\ket{\phi}\!\bra{\phi})= \tau_B$ and $\tr_B(\ket{\phi}\!\bra{\phi})= \tau_A$.
Note that, in the limit $T\to\infty$, the Gibbs states are maximally mixed, and hence the $\vr_{\alpha}$ are isotropic states.

\begin{figure}
    \centering
    \includegraphics[width=1.\columnwidth]{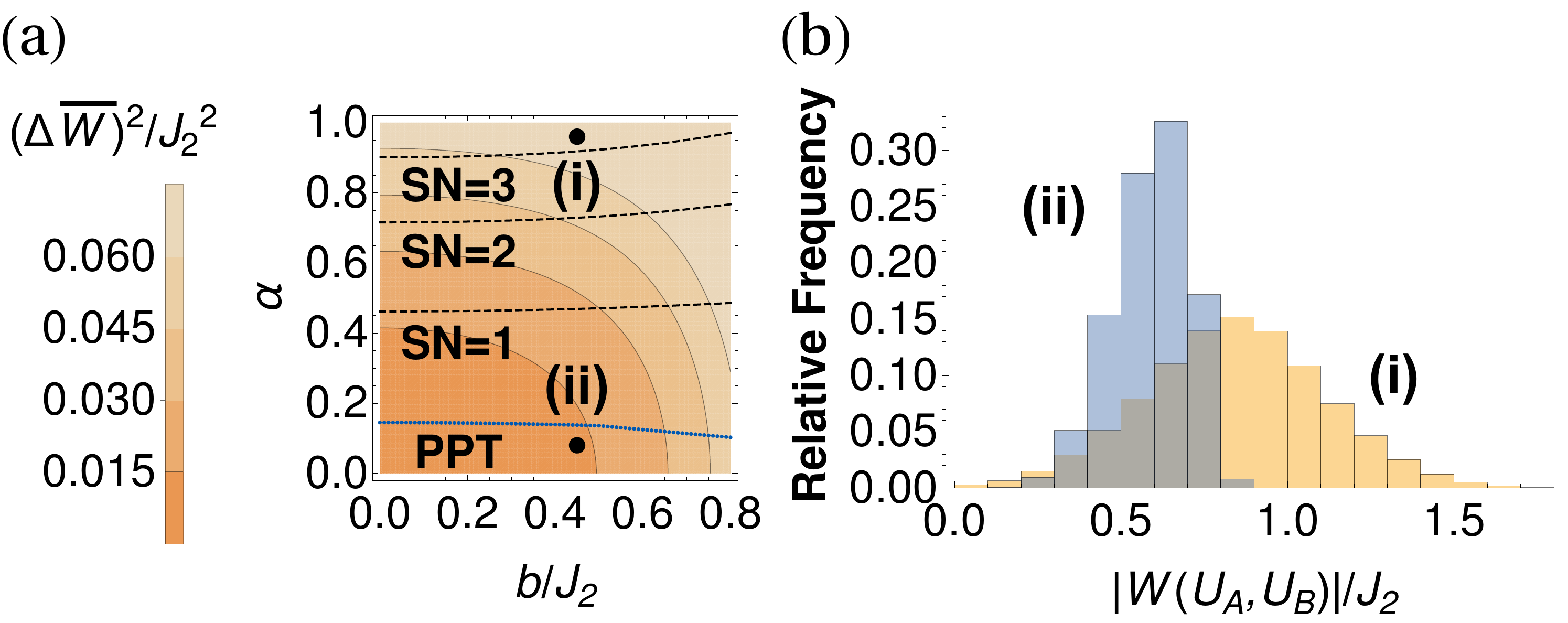}
    \caption{Schmidt number detection through local work fluctuations in an Ising-type battery of $2+2$ qubits. (a) Variance of average work extracted by local random unitaries acting on each battery half as a function of the field strength $b$ and the mixing ratio $\alpha$ between a maximally entangled and a product Gibbs state. 
    All energies are in units of the interaction strength $J_2$, and we fix $J_{1,3} = 0.5 J_2$ and $T=1.5 J_2$. Quantum states with $\rm{SN}=1,2,3$ are contained in the areas below the respective dashed lines, according to Eq.~\eqref{SNdetect_from_WFs}, so above a line allows us to detect SN.
    For comparison, we also indicate a bottom blue threshold given by the PPT criterion.
    (b) Exemplary histograms of negative work values from a sample of $10^6$ unitaries for the two marked cases (i) and (ii) at $b=0.45$, corresponding to an entangled state of $\rm{SN}=4$ at $\alpha=0.96$ and a state at $\alpha=0.08$, compatible with separable states, respectively. Work values are divided into bins of size $0.1 J_2$.
    }
    \label{fig:4by4alpha}
\end{figure}

As a simple example, consider an interacting four-qubit battery based on the Ising-type Hamiltonian 
\begin{align}
    H_{I} &= \sum_{i=1,2,3} J_{i} Z_i \otimes Z_{i+1}
    + b \sum_{i=1}^4 Z_{i},
    \label{eq-hamiltonian}
\end{align}
with $Z_{i}$ being the Pauli-$Z$ matrix acting on the $i$-th qubit, $b$ the homogeneous field strength, and $J_i$ the nearest-neighbour couplings.
Assuming the bipartition $(A|B)=(1,2|3,4)$, we can identify $h_A^2 =J_1^2+2b^2$,
$h_B^2 =J_3^2+2b^2$, and
$g^2v^2=J_2^2$. We illustrate the work fluctuations for an exemplary choice of strong coupling parameters in Fig.~\ref{fig:4by4alpha}.
Panel (a) shows the work variance as a function of $(b,\alpha)$
and the Schmidt-number thresholds for $k=1,2,3$, while (b) shows two selected histograms of suitably binned work values Eq.~\eqref{extractablework} associated with the Haar-random local unitaries. 
In practice, these values could be inferred from joint local measurements in the $Z$-basis on sufficiently many identical copies of each unitary sample, and the statistical significance can be evaluated according to Ref.~\cite{ketterer2020certifying}. In the following, however, we will 
proceed to introduce two different measurement schemes to estimate the work fluctuations.

\section{Noisy two-point energy measurement protocol}
The projective two-point measurement (TPM) protocol \cite{talkner2007fluctuation,esposito2009nonequilibrium,campisi2011colloquium} defines a quantum notion of fluctuating work in analogy to classical stochastic thermodynamics, for trajectories of an arbitrary system state subject to a given isentropic process $U$. 
In this protocol, one first performs a projective measurement in the system's energy eigenbasis, lets the post-measurement state evolve under $U$, and then performs a second projective energy measurement. The difference between both outcomes can be seen as a random realization of work under $U$, and the so defined work statistics obey the Jarzynski equality~\cite{talkner2007fluctuation,esposito2009nonequilibrium,campisi2011colloquium}. 

However, the protocol has two major downsides. First, it is highly invasive since the first measurement voids all the coherence between energy levels that the initial system state might have. Genuine quantum signatures such as entanglement between different system parts may thus be destroyed. 
Second, ideal projective measurements may not be achievable due to limited accuracy and unavoidable noise in experiments. These two problems have motivated recent efforts to generalize the TPM protocol
\cite{roncaglia2014work,de2015measuring,talkner2016aspects,perarnau2017no,baumer2018fluctuating,de2018ancilla,lostaglio2018quantum,bednorz2010quasiprobabilistic,guryanova2020ideal, debarba2019work, beyer2021joint}.

We alleviate both problems by employing a TPM protocol with noisy detectors, first introduced in Ref.~\cite{beyer2021joint}. We adapt it to our setting of composite quantum batteries 
and have $A, B$ each apply the protocol for a \textit{local} energy measurement.
To this end, we expand $H_{X}=\sum_{i=1}^{d} E_i^X \Pi_i^X$, with the energy eigenvalues $E_i$ and the projectors $\Pi_i^X$ to the corresponding eigenspaces. Moreover, we write the interaction term as $V = \sum_{i,j=1}^d D_{ij} \Pi_i^A \otimes \Pi_j^B + V_{od}$ with $\tr[V_{od}\Pi_i^A \otimes \Pi_j^B]=0$ for all $i,j$. This separates mere level shifts of the joint diagonal energy spectrum, $E_{ij} = E_i^A + E_j^B + gD_{ij}$, from the actual change of the energy eigenbasis via the off-diagonal part $V_{od}$. 
The following results are based on estimating the $E_{ij}$-spectrum from noisy measurements in the basis of the $\Pi_i^A \otimes \Pi_j^B$. We stress that, for $V_{od} \neq 0$, the $E_{ij}$-values are not the battery energies and the measurement does not constitute an actual energy measurement (though it approximates one for small $V_{od}$).

The population of the diagonal spectrum $(E_{ij})$ 
can be probed straightforwardly by combining the outcomes of local energy measurements. Suppose these measurements are erroneous in that they detect the correct local energy state only with probability $\varepsilon$, while producing a completely random outcome with probability $1-\varepsilon$. Assuming the same $\varepsilon$ for both sides, the corresponding POVMs are
\begin{equation}
    P_i^X = \varepsilon \Pi_i^X + \frac{1-\varepsilon}{d}\eins_X, \quad \sum_{i=1}^d P_i^X = \eins_X.
    \label{eq-povm-tpm}
\end{equation}
Here we assume that the $\Pi_i^X$ are rank-$1$ projectors, so that the entire POVM has $d$ outcomes.
On average, we can obtain an unbiased estimator for $(E_{ij})$
from them by assigning to each joint outcome $(ij)$ occuring with probability $m_{ij} = \tr[P_i^A\otimes P_j^B \vr_{AB}]$ the rescaled and shifted energy value \cite{beyer2021joint}
\begin{equation}
    e_{ij} = \frac{E_i^A + E_j^B}{\varepsilon} + \frac{gD_{ij}}{\varepsilon^2} - \frac{1-\varepsilon}{d\varepsilon} \left( \tr[H_A] + \tr[H_B] \right).
\end{equation}
For $\varepsilon=1$, we have noiseless projective measurements and $e_{ij}=E_{ij}$, whereas small values $\varepsilon\ll 1$ correspond to a weak measurement dominated by errors.
{Note that the case of different errors $\varepsilon_A, \varepsilon_B$ is also discussed in Appendix \ref{app:B}.}

We subject the post-measurement state to local random unitaries $U_A\otimes U_B$,
\begin{align}
    \sigma_{ij} = \frac{
    U_A\sqrt{P^A_i}\otimes U_B\sqrt{P^B_j} \vr_{AB}
    \sqrt{P^A_i}U_A^\dagger \otimes \sqrt{P^B_j}U_B^\dagger}{m_{ij}},
\end{align}
before applying the same local measurement again.
The probability to obtain $e'_{kl}$ if the first outcome was $e_{ij}$ is $m_{kl|ij} = \tr[P^A_k\otimes P^B_l \sigma_{ij}]$, to which we associate a presumed work value $w_{ijkl} = e_{ij}-e_{kl}^\prime$. (It may only approximate the extracted work if $V_{od}\neq 0$, but small.) 
Averaged over many repetitions at fixed $U_A\otimes U_B$, we define $W_{{\rm TPM}} (\varepsilon) \equiv W_{{\rm TPM}} (\varepsilon, U_A,U_B) = \sum_{i,j,k,l} m_{ij} m_{kl|ij} w_{ijkl}$, which in turn can be averaged over a large sample of unitaries to yield
\begin{align}\label{NTPMworkmeasuremeny}
    \overline{W_{{\rm TPM}} (\varepsilon)}
    &=\int dU_A \int dU_B  \,
    W_{{\rm TPM}} (\varepsilon),\\
    \va{\overline{W_{{\rm TPM}} (\varepsilon)}} &=
    \overline{W_{{\rm TPM}} (\varepsilon)^2}
    -\overline{W_{{\rm TPM}} (\varepsilon)}^2. \label{eq:NTPMworkvar}
\end{align}
In general, these TPM cumulants do not coincide with the previously
defined ones in Eqs.~\eqref{eq:workMean} and \eqref{eq:workVariance}. 
However, we can still obtain an explicit relation between the variances:

\textbf{Observation 3.}
\textit{
For any $d \times d$ composite quantum battery described by $\vr_{AB}$ 
and $H_{AB}$, the local noisy TPM protocol results in the presumed work variance
\begin{align}
    \va{\overline{W_{{\rm TPM}} (\varepsilon)}}
    &= n_0(\varepsilon)
    \va{\overline{W}}_{D} + 
    n_1(\varepsilon)
    \va{\overline{W_{{\rm Proj}}}} \nonumber \\
    &\quad + [1-n_0(\varepsilon) - n_1(\varepsilon)]\va{\overline{W_{{\rm Noisy}}}}, \label{eq:varW_TPM}
\end{align}
where the functions $n_{0,1}(\varepsilon) \in [0,1]$ for any $\varepsilon \in [0,1]$ are explicitly given. 
The term $\va{\overline{W}}_{D}$ is the theoretical work variance in Eq.~\eqref{eq:varW_ideal} evaluated for $V_{od}=0$. The $\va{\overline{W_{{\rm Proj}}}}$ and $\va{\overline{W_{{\rm Noisy}}}}$ represent the variance for a noiseless projective TPM and an additional contribution at finite noise $\varepsilon \in (0,1)$, respectively, both also at $V_{od}=0$.
}

\begin{figure}[t]
    \centering
    \includegraphics[width=1.\linewidth]{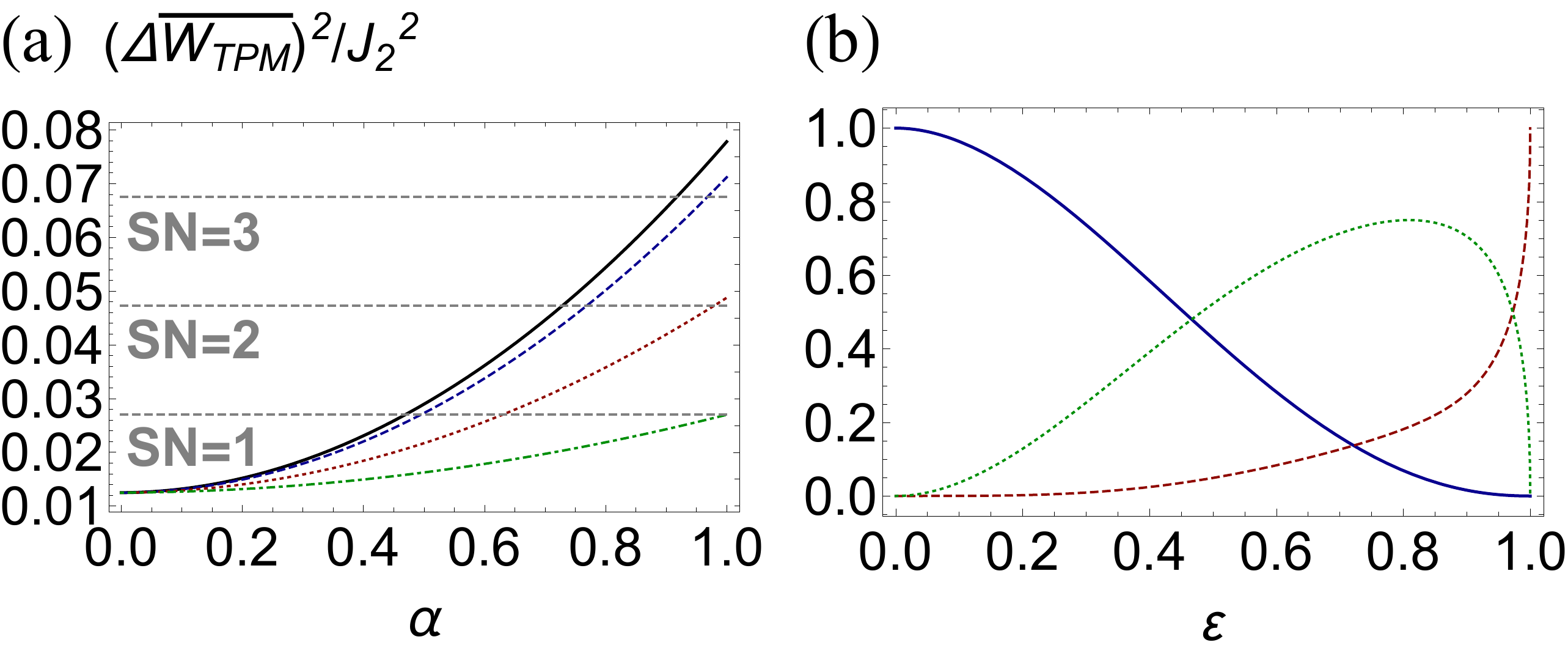}
    \caption{
    (a) Comparison between the theoretical work variance $\va{\overline{W}}_{D}$ (black solid) and the variance $\va{\overline{W_{{\rm TPM}} (\varepsilon)}}$ resulting from a local TPM protocol at various noise levels
    $\varepsilon = 0.2, 0.5,$ and $1.0$ (respectively, dashed blue, dotted red, and dash-dotted green), for the Ising battery of Fig.~\ref{fig:4by4alpha} at fixed $b=0.45J_2$ and varying mixing ratio $\alpha$. The dashed horizontal lines show the bounds compatible with Schmidt numbers $1,2,3$. 
    (b) Weight functions
    $n_0(\varepsilon)$ (blue solid),
    $n_1(\varepsilon)$ (dashed red), and
    $1-n_0(\varepsilon)-n_1(\varepsilon)$ (dotted green) versus noise level $\varepsilon$}.
    \label{NTPEPMP}
\end{figure}

{See Observation 6 in Appendix \ref{app:B}} for the proof and the lengthy explicit expressions for 
$\va{\overline{W_{{\rm Proj}}}}$, $\va{\overline{W_{{\rm Noisy}}}}$, and $n_{0,1} (\varepsilon)$.
There we also show that the noisy TPM variance obeys $\va{\overline{W_{{\rm TPM}} (\varepsilon)}} \leq \va{\overline{W}}_{D}$, which saturates in the limit $\varepsilon \to 0$, where $n_0 \to 1$ and $n_1 \to 0$.
In the opposite limit $\varepsilon \to 1$ where $n_1 \to 1$ and $n_0 \to 0$, we have a local projective TPM which does not detect any entanglement.
We compare the measured work variance at various noise levels to the theoretical values for our example states Eq.~\eqref{eq:thermalWstates} in Fig.~\ref{NTPEPMP}, demonstrating that the noisy local TPM can detect entanglement.

\section{Noisy energy coincidence measurement protocol}
In order to estimate the work variance in Eq.~\eqref{eq:NTPMworkvar}, the noisy 
TPM scheme still relies on subjecting many copies of the battery state to the 
same randomly drawn local unitary. We can reduce this overhead by performing local 
coincidence measurements on merely two state copies $\vr_{AB} \otimes \vr_{A^\prime B^\prime}$ subjected to the same local unitary $U_A \otimes U_B$. 
Ideally, a joint dichotomic projective measurement $\Pi_{AA^\prime} \otimes \Pi_{BB^\prime}$ would act  locally on both $A$-copies and on both $B$-copies, with 
$\Pi_{XX^\prime} = \sum_{i} \Pi_{i}^{X} \otimes \Pi_{i}^{X^\prime}$ projecting 
onto the subspace spanned by energy product states with the same eigenvalues 
$E_i^X = E_i^{X'}$. By repeating this measurement with a large sample of 
Haar-random unitaries, we could estimate the average probability 
$\overline{\mathcal{C}}$ that the two copies' local energies on the $A$- and on the $B$-side 
both coincide. 

More generally, we can define a dichotomic energy coincidence POVM based on noisy local energy measurements according to Eq.~\eqref{eq-povm-tpm},
\begin{equation}
    P_{XX'} = \sum_i P_i^X \otimes P_i^{X'} = \varepsilon^2 \Pi_{XX'} + \frac{1-\varepsilon^2}{d} \eins_{XX'}
\end{equation}
The probability for local energy coincidence between the copies on both sides is then  $\mathcal{C}(\varepsilon) = \tr [P_{AA'}\otimes P_{BB'} \vr_{AB}' \otimes \vr_{A'B'}']$. 
Averaged over the unitaries,
\begin{equation}
    \overline{\mathcal{C}(\varepsilon)}
    = \frac{1}{d^2}
    \left[1+\frac{(r_A^2+r_B^2)\varepsilon^2}{d+1} +\frac{t^2\varepsilon^4}{(d+1)^2} \right],
    \label{noisyenercoinmeas}
\end{equation}
which we can directly relate to the entanglement-sensitive work variance 
$\va{\overline{W}}$ from Eq.~(\ref{eq:workVariance}).
{Eq.~(\ref{noisyenercoinmeas}) can be derived more generally, using different errors $\varepsilon_A, \varepsilon_B$ for measurements on the $A,B$ sides:

\begin{proof}
First, we can immediately find
\begin{align} \nonumber
    \overline{\mathcal{C}(\varepsilon_A,\varepsilon_B)}
    =\tr
    \left[
    \Phi(\vr_{AB})
     P_{AA^\prime} \otimes P_{BB^\prime}
    \right],
\end{align}
where $\Phi(\vr_{AB})$ is defined in Eq.~(\ref{PhivrABmap}).
Since
\begin{align}
    \tr[P_{XX^\prime}]
    &=\varepsilon_X^2 \tr[\Pi_{XX^\prime}] + \frac{1-\varepsilon_X^2}{d}\tr[\eins_{XX^\prime}]
    =d,\\
    \tr[S_{X} P_{XX^\prime}] \nonumber
    &= \varepsilon_X^2 \tr[S_X\Pi_{XX^\prime}] + \frac{1-\varepsilon_X^2}{d}\tr[S_X\eins_{XX^\prime}]\\
    &=(d-1)\varepsilon_X^2+1,
\end{align}
we find
\begin{align} \nonumber
    \overline{\mathcal{C}(\varepsilon_A,\varepsilon_B)}
    =\frac{1}{d^2}
    \left[
    1
    + \frac{r_A^2\varepsilon_A^2}{d+1}
    + \frac{r_B^2\varepsilon_B^2}{d+1} 
    + \frac{t^2\varepsilon_A^2\varepsilon_B^2}{(d+1)^2}
    \right].
\end{align}
For $\varepsilon_A=\varepsilon_B=\varepsilon$,
we arrive at Eq.~(\ref{noisyenercoinmeas}).
\end{proof}
}

Expressing the battery interaction strength as $g^2v^2=(d-1)(h^2\varepsilon^2+c)$, with $h^2 = \min{(h_A^2,h_B^2)}$ and a new term $c$, we find:

\textbf{Observation 4.}
\textit{
In the noisy energy coincidence measurement protocol, we have 
\begin{equation}
    \overline{\mathcal{C}(\varepsilon)}
    \leq \frac{1}{d^2} \left[1+ \frac{(d-1)\varepsilon^2}{h^2}\va{\overline{W}}
    + \frac{t^2 \varepsilon^2 (|c|-c)}{2(d+1)^2 h^2} \right].
\end{equation}
}

Hence, the energy coincidence measurement protocol on two identical copies gives access to nonlinear functions of the battery state such as the work variance, which allows us to detect the Schmidt number by virtue of Observation 2. 

The proven influence of the Schmidt number on work fluctuations exemplifies the observable thermodynamic implications of high-dimensional bipartite entanglement. Our assessment in terms of the work variance with respect to Haar-random samples of unitaries extends previous studies on the direct estimation of nonlinear functions \cite{horodecki2002method,horodecki2003measuring,ekert2002direct,bovino2005direct,carteret2005noiseless,schmid2008experimental}, experimental lower bounds on the concurrence \cite{mintert2005concurrence,aolita2006measuring,mintert2007observable,walborn2007experimental}, and protocols for randomized measurements \cite{van2012measuring,elben2019statistical,elben2020cross}.

\section{Conclusion}
We have analyzed the role of entanglement in local work exchange with a composite quantum battery, as described by an interacting bipartite quantum system. Specifically, we found that the variance of the average extracted work over a Haar-random sample of local unitary processes obeys a hierarchy of inequalities detecting the Schmidt number of the battery state---a criterion for high-dimensional entanglement. While we saw that these bounds cannot be probed directly in a standard projective two-point measurement scheme, we could show that Schmidt number detection is possible in a two-point measurement with noisy detectors as well as in an energy coincidence measurement.

Our results {can be used to probe the influence of entanglement in a quantum working medium and, more generally,} elucidate the interplay between entanglement and energy fluctuations in random processes. 
{It would be interesting to verify our results on experimental platforms for quantum thermal machines and batteries. Moreover, the randomized two-point measurement approach could be extended to non-unitary, dissipative processes, facilitating the detection of heat leaks and non-unital dynamics in complex open quantum systems \cite{kafri2012holevo, gardas2018quantum}.}

\begin{acknowledgments}
We would like to thank
Roope Uola,
Zhen-Peng Xu,
and
Xiao-Dong Yu
for discussions.
This work was supported by
the Deutsche Forschungsgemeinschaft (DFG, German Research Foundation, 
Projects No. 447948357 and No. 440958198),
the Sino-German Center for Research Promotion (Project M-0294),
the ERC (Consolidator Grant No. 683107/TempoQ),
the German Ministry of Education and Research (Project QuKuK, BMBF 
Grant No. 16KIS1618K) and the DAAD.
\end{acknowledgments}

\clearpage
\onecolumngrid
\appendix

\section{Useful formulas}\label{app:A}
Here we summarize useful formulas related to the SWAP operation and to integrals over Haar-random unitaries.
\begin{itemize}
    \item
    For all operators $A$ and $B$,
    $\tr[A\otimes B] = \tr[A] \cdot \tr[B]$
    and
    $\tr[A^{\otimes k}] = \tr[A]^{k}$.
    
    \item
    Let $S$ be the SWAP (flip) operator acting on
    $d \times d$-dimensional systems, defined as
    $S\ket{a}\ket{b}= \ket{b}\ket{a}$.
    The SWAP operator $S$ can be written as
    $S=d\ket{\Psi^+}\!\bra{\Psi^+}^{T_A}$,
    where 
    $\ket{\Psi^+} = (1/\sqrt{d}) \sum_{i=0}^{d-1}\ket{ii}$
    is the maximally entangled state
    and $T_A$ is the partial transposition on $A$.
    That is, 
    $S=\sum_{i,j=0}^{d-1} \ket{ij}\!\bra{ji}$
    with $\pm 1$ eigenvalues.
    Another and useful expression of the SWAP operator is given by
    \begin{align}
        S = \frac{1}{d} \sum_{i=0}^{d^2-1} \lambda_i \otimes \lambda_i.
    \end{align}    
    For the two-qubits Bell state
\begin{align}\label{twoqubitbell}
    \ket{{\Phi^+}}\!\bra{{\Phi^+}} =\frac{1}{4}
\left(\eins_2 \otimes \eins_2
+ X \otimes X
- Y \otimes Y
+ Z \otimes Z
\right),
\end{align}
where
$\ket{{\Phi^+}}=(\ket{00}+\ket{11})/\sqrt{2}$
and
with $X,Y,Z$ the Pauli matrices,
this can be seen since Pauli $Y$ satisfies $(Y\otimes Y)^{T_A} =-Y\otimes Y$.
    Here we note the important property:
    \begin{align}
    \label{swapexchange}
    \tr[S (A\otimes B)] = \tr[AB].
    \end{align}
    
    \item
    Let $dU$ be the Haar measure, that is, the uniformly random measure on the group of unitary operations $\mathcal{U}(d)$, normalized to $\int dU = 1$.
    For any integrand function $f(U)$ on $U\in \mathcal{U}(d)$, the Haar measure is both left and right invariant under shifts by any unitary operation $V \in \mathcal{U}(d)$,
    \begin{align}
        \label{leftrightinvariant}
        \int dU \, f(U) &= \int dU \, f(V U) = \int dU \, f(UV).
    \end{align}

    \item
    For an operator $X \in \left(\mathcal{H}^D\right)^{\otimes k}$,
    let us consider
    \begin{align}
    \Lambda_{k}(X)= \int dU \, U^{\otimes k} X (U^\dagger)^{\otimes k}.
    \end{align}
    It is known that the unitary integral can be evaluated using the Schur-Weyl duality and the Weingarten calculus,
    see Refs.~\cite{goodman2000representations, roberts2017chaos, kliesch2021theory}.
    In the cases of $k=1, 2$, they are given by
    \begin{align}
    \label{Phi1map}
    \Lambda_{1}(X) &= \int dU\, U X U^\dagger
    =\frac{\tr[X]}{D} \eins_D,\\
    \label{Phi2map}
    \Lambda_{2}(X) &= \int dU\, U^{\otimes 2} X (U^\dagger)^{\otimes 2}
    =\frac{1}{D^2-1}
    \left\{
    \left[\tr(X)-\frac{\tr(SX)}{D}\right] \eins_D^{\otimes 2}
    +\left[\tr(SX)-\frac{\tr(X)}{D}
    \right]S
    \right\}.
    \end{align}
    One lesson from this result is that $\Lambda_2(X)$ realizes a SWAP operation.
    That is, taking integrals over the Haar unitary for second moments can yield an indirect application of the SWAP operation.
    
    \item
    For a two-qudit state $\vr_{AB}$, let us consider 
    \begin{align}
    \Phi(\vr_{AB})=\int dU_A dU_B \,
    (U_A^{\otimes 2} \otimes U_B^{\otimes 2})\vr_{AB}^{\otimes 2}(U_A^\dagger)^{\otimes 2} \otimes (U_B^\dagger)^{\otimes 2}.
    \end{align}
    Using the generalized Bloch representation of $\vr_{AB}$ and the above formulas, we can obtain
\begin{align}
\label{PhivrABmap}
\nonumber
    \Phi(\vr_{AB})
    &= \frac{1}{d^4} \Bigg\{
    \eins_{AB}^{\otimes 2}
    +\frac{1}{d^2-1}
    \Bigg[
    r_A^2\left(dS_{A} - \eins_{A}^{\otimes 2}  \right) \otimes \eins_{B}^{\otimes 2}
    +r_B^2\eins_{A}^{\otimes 2} \otimes \left(dS_{B} - \eins_{B}^{\otimes 2}
    \right)\\
    &\quad
    +\frac{t^2\left( dS_{A}-\eins_{A}^{\otimes 2} \right)
    \otimes \left( dS_{B}-\eins_{B}^{\otimes 2} \right)}{d^2-1}
    \Bigg]
    \Bigg\},
\end{align}
where
$S_A$ and $S_B$ respectively are the SWAP operators acting on the two-copy system of $\vr_{AB}^{\otimes 2}$.
\end{itemize}

\section{Noisy two-point energy measurement protocol}\label{app:B}
\subsection{Noisy energy measurement and general observations}
We consider noisy local energy measurements on $A$ and $B$ with errors $\varepsilon_A$, $\varepsilon_B$,
\begin{align}
    P_i^A = \varepsilon_A \Pi_i^A + \frac{1-\varepsilon_A}{d}\eins_A,
    \,\,\,\,\,\,\,\,\,\,\,\,~
    P_i^B = \varepsilon_B \Pi_i^B + \frac{1-\varepsilon_B}{d}\eins_B.
\end{align}
In the main text, we assumed $\varepsilon_A = \varepsilon_B$. The probability to obtain the local measurement outcomes $i,j$ on $\vr_{AB}$ is given by
$
    m_{ij} = \tr[P_i^A\otimes P_j^B \vr_{AB}].
$
Following the notion of quantum instruments \cite{heinosaari2011mathematical}, the normalized post-measurement state can be described by
\begin{align} 
    \sigma_{ij} = \frac{1}{\tr[\mathcal{J}_{ij}(\vr_{AB})]}
    \mathcal{J}_{ij}(\vr_{AB}),
\end{align}
where $\mathcal{J}_{ij}$ is a linear completely positive and trace-preserving (CPTP) map satisfying
\begin{align} \label{instrumentproperty}
     m_{ij} = \tr[\mathcal{J}_{ij}(\vr_{AB})].
\end{align}
Like most studies on two-point measurement protocols, we employ the so-called von Neumann-L\"uders instrument in the main text,
\begin{align}
    \mathcal{J}_{ij}^{\text{vN-L}}(\vr_{AB})
=\sqrt{P_i^A}\otimes \sqrt{P_j^B} \vr_{AB}
    \sqrt{P_i^A}\otimes \sqrt{P_j^B}.
\end{align}

For the sake of simplicity, let us now define the diagonal Hamiltonian $H_D$ as an effective description:
\begin{align}
    H_D = H_{AB} - gV_{od},
\end{align}
where $V = \sum_{i,j=1}^d D_{ij} \Pi_i^A \otimes \Pi_j^B + V_{od}$
and $V_{od}$ is the off-diagonal part of the interaction Hamiltonian (with vanishing diagonal elements) in the eigen-energy basis of the local Hamiltonian.
On the one hand, the Hamiltonian $H_D$ can be decomposed using the corresponding projectors $\Pi_i^A \otimes \Pi_j^B$,
\begin{align}
    H_D = \sum_{i,j} E_{ij} \Pi_i^A \otimes \Pi_j^B,
\end{align}
with the joint diagonal energy spectrum
$E_{ij} = E_i^A + E_j^B + gD_{ij}$, given in the main text.
On the other hand, the Hamiltonian $H_D$ can also be decomposed into the measurement operators $P_i^A, P_j^B$,
\begin{align}
   H_D = \sum_{i,j}e_{ij} P_i^A \otimes P_j^B,
\end{align}
with appropriate energy values $e_{ij}$ assigned to each pair of measurement outcomes $(i,j)$,
\begin{align}
    e_{ij} &= e_i^A + e_j^B + gd_{ij},\\
    e_i^A  &= \frac{1}{\varepsilon_A} E_i^A - \frac{1-\varepsilon_A}{d\varepsilon_A}\tr[H_A],\,\,\,\,\,\,~
    e_j^B  = \frac{1}{\varepsilon_B} E_j^B - \frac{1-\varepsilon_B}{d\varepsilon_B}\tr[H_B],\,\,\,\,\,\,~
    d_{ij} = \frac{1}{\varepsilon_A \varepsilon_B} D_{ij}.
\end{align}
The POVM decomposition of the Hamiltonian is motivated by the research in Ref.~\cite{beyer2021joint}.
For $\varepsilon_A, \varepsilon_B=1$, we have noiseless projective measurements and $e_{ij}=E_{ij}$, whereas small values $\varepsilon_A, \varepsilon_B \ll 1$ correspond to a weak measurement dominated by errors.

Similarly with the main text, we define the average work over the noisy TPM protocol for independent errors as
\begin{align}
    W_{{\rm TPM}} (\varepsilon_A, \varepsilon_B)
    \equiv
    W_{{\rm TPM}}(\varepsilon_A, \varepsilon_B, U_A, U_B)
    = \sum_{i,j,k,l} m_{ij} m_{kl|ij} w_{ijkl}.
\end{align}
Here we recall that
$    m_{kl|ij} = \tr[P_k^A\otimes P_l^B \sigma_{ij}^\prime]$
is the conditional probability to obtain the outcomes $k,l$ associated to the energy value $e_{kl}^\prime$ in the second measurement, given that we obtained $(i,j)$ and $e_{ij}$ in the first measurement. The second measurement receives the state $\sigma_{ij}^\prime = (U_A\otimes U_B) \sigma_{ij} (U_A\otimes U_B)^\dagger$, which is the state transformed by a local random unitary operation after the first noisy energy measurement. 
We associate the presumed work value $w_{ijkl} = e_{ij}-e_{kl}^\prime$ to the outcomes. 
Taking an average over a large sample of local unitaries yields
\begin{align}
    \overline{W_{{\rm TPM}} (\varepsilon_A, \varepsilon_B)}
    &=\int dU_A \int dU_B  \,
    W_{{\rm TPM}} (\varepsilon_A, \varepsilon_B),\\
    \va{\overline{W_{{\rm TPM}} (\varepsilon_A, \varepsilon_B)}} &=
    \overline{W_{{\rm TPM}} (\varepsilon_A, \varepsilon_B)^2}
    -\overline{W_{{\rm TPM}} (\varepsilon_A, \varepsilon_B)}^2.
\end{align}

In the following, we evaluate and simplify the unitary integrals:

\textbf{Observation 5.}
\textit{
For any $d \times d$ composite quantum battery described by $\vr_{AB}$ 
and $H_{D}$, the local noisy TPM protocol with $\varepsilon_A$ and $\varepsilon_B$ for the von Neumann-L\"uders instrument results in the average which can 
be expressed as
\begin{align}
    \overline{W_{{\rm TPM}} (\varepsilon_A, \varepsilon_B)}
    = \tr[\vr_{AB} H_{D}] - \frac{\tr[H_{D}]}{d^2}.
\end{align}
}

\textbf{Observation 6.}
\textit{
For any $d \times d$ composite quantum battery described by $\vr_{AB}$ 
and $H_{D}$ with $\tr[H_{D}]=0$, the local noisy TPM protocol with $\varepsilon_A$ and $\varepsilon_B$ for the von Neumann-L\"uders instrument results in the presumed work variance which can 
be expressed as
\begin{align}
    \va{\overline{W_{{\rm TPM}} (\varepsilon_A, \varepsilon_B)}}
    =\Upsilon_{{\rm Ideal}}
    +\Upsilon_{{\rm Proj}}
    +\Upsilon_{{\rm Noisy}},
\end{align}
where
$\Upsilon_{{\rm Ideal}}$,
$\Upsilon_{{\rm Proj}}$, and
$\Upsilon_{{\rm Noisy}}$,
respectively,
represent the effects of
the ideal theoretical work variance,
the variance from a noiseless projective TPM, and
the noisy additional measurements at finite noise.
They are given by
\begin{align}
    \Upsilon_{{\rm Ideal}}
    &\equiv
    \kappa_{AB}^2 \va{\overline{W}}_{D},
    \\ \nonumber
    \Upsilon_{{\rm Proj}}
    &\equiv
    \frac{1}{d^2-1}\Bigg\{
    \Big[
    (f_{\varepsilon_A}^4 f_{\varepsilon_B}^4+\kappa_A^2)
    \left(dp_A^2-1\right)+\kappa_B^2r_A^2
    \Big]h_A^2
    +\Big[
    (f_{\varepsilon_A}^4 f_{\varepsilon_B}^4+\kappa_B^2)
    \left(dp_B^2-1\right)+\kappa_A^2r_B^2
    \Big]h_B^2\\
    &\quad
    +\frac{g^2v^2}{d^2-1}
    \Big[
    f_{\varepsilon_A}^4 f_{\varepsilon_B}^4 \left(d^2p_{AB}^2 -dp_A^2 - dp_B^2 +1 \right)
    +\kappa_A^2 \sum_{a,b,c}t_{ab}t_{cb}\zeta_{ac}^A
    +\kappa_B^2 \sum_{a,b,c}t_{ab}t_{ac}\zeta_{bc}^B
    \Big]
    \Bigg\},
    \\ \nonumber
    \Upsilon_{{\rm Noisy}} 
    &\equiv
    \frac{2}{d^2-1}\Bigg\{
     \Big[\gamma_A\left(dp_A^2-1\right) +\kappa_B\kappa_{AB} r_A^2\Big]h_A^2
    +\Big[\gamma_B\left(dp_B^2-1\right) +\kappa_A\kappa_{AB} r_B^2\Big]h_B^2\\
    &\quad
    +
    \frac{g^2v^2}{d^2-1}
    \Big[
    \gamma_{AB}\left(d^2p_{AB}^2 -dp_A^2 - dp_B^2 +1 \right)
    \Big]
    +\kappa_A \kappa_{AB} \sum_{a,b,c}t_{ab}t_{cb}\zeta_{ac}^A
    +\kappa_B \kappa_{AB} \sum_{a,b,c}t_{ab}t_{ac}\zeta_{bc}^B
    \Bigg\}.
\end{align}
Here, 
$\va{\overline{W}}_{D}$ is the ideal theoretical work variance, Eq.~\eqref{eq:varW_ideal} in the main text, evaluated for the diagonal Hamiltonian $H_D$ (i.e., for $V_{od}=0$ and $v^2=(1/d^2)\sum_{i,j}D_{ij}^2$).
In the above expressions, we introduce the short-hand notations
\begin{align}
    \gamma_A
    &\equiv
    f_{\varepsilon_A}^2 f_{\varepsilon_B}^2
    (\kappa_A + \kappa_B + \kappa_{AB})
    +\kappa_A( \kappa_B +\kappa_{AB}),\\
    \gamma_B
    &\equiv
    f_{\varepsilon_A}^2 f_{\varepsilon_B}^2
    (\kappa_A + \kappa_B + \kappa_{AB})
    +\kappa_B( \kappa_A +\kappa_{AB}),\\
     \gamma_{AB}
     &\equiv
     f_{\varepsilon_A}^2 f_{\varepsilon_B}^2
    (\kappa_A + \kappa_B + \kappa_{AB})+\kappa_A \kappa_B,\\
    \kappa_{AB}
    &\equiv {\kappa_A \kappa_B}/({f_{\varepsilon_A}^2 f_{\varepsilon_B}^2}),\ \ \ \ 
    \kappa_A
    \equiv
    f_{\varepsilon_A}^2 g_{\varepsilon_B}
    \left(2f_{\varepsilon_B}+dg_{\varepsilon_B}
    \right),\ \ \ \ 
    \kappa_B
    \equiv
    f_{\varepsilon_B}^2 g_{\varepsilon_A}
    \left(2f_{\varepsilon_A}+dg_{\varepsilon_A}
    \right),\\
    f_{\varepsilon_X}
    &\equiv \sqrt{\varepsilon_X + \frac{1-\varepsilon_X}{d}} -\sqrt{\frac{1-\varepsilon_X}{d}},\\
    g_{\varepsilon_X}
    &\equiv \sqrt{\frac{1-\varepsilon_X}{d}},\\
    p_{AB}^2
    &\equiv \sum_{i,j} (p_{ij}^{AB})^2,\ \ \ \
    p_A^2
    \equiv \sum_{i} (p_{i}^A)^2,\ \ \ \
    p_B^2
    \equiv \sum_{j} (p_{j}^B)^2,\\
    p_{ij}^{AB}
    &\equiv \tr[\Pi_{i}^{A}\otimes \Pi_j^B \vr_{AB}],\ \ \ \
    p_{i}^A
    \equiv \sum_{j} p_{ij}^{AB},\ \ \ \
    p_{j}^B
    \equiv \sum_{i} p_{ij}^{AB},\\
    \zeta_{ab}^A
     &\equiv
     \sum_i \frac{\tr(\Pi_{i}^{A} \lambda_a \Pi_{i}^{A} \lambda_b )}{d},
     \ \ \ \ 
     \zeta_{ab}^B
     \equiv
     \sum_i \frac{\tr(\Pi_{i}^{B} \lambda_a \Pi_{i}^{B} \lambda_b )}{d},
\end{align}
with the normalization condition
\begin{align}
    f_{\varepsilon_X}^2+2f_{\varepsilon_X}g_{\varepsilon_X}+dg_{\varepsilon_X}^2=1,
\end{align}
for $X=A,B$.
Let us define
\begin{align}
    n_{0}(\varepsilon_A, \varepsilon_B)
    &\equiv
    \kappa_{AB}^2,\\
    n_{1}(\varepsilon_A, \varepsilon_B)
    &\equiv
    f_{\varepsilon_A}^4 f_{\varepsilon_B}^4+\kappa_A^2+\kappa_B^2
    ,\\
    n_{{\rm Noisy}}(\varepsilon_A, \varepsilon_B)
    &\equiv
    2\left[f_{\varepsilon_A}^2 f_{\varepsilon_B}^2
    (\kappa_A + \kappa_B + \kappa_{AB})
    +\kappa_A \kappa_B+\kappa_A \kappa_{AB}+\kappa_B \kappa_{AB}
    \right],
\end{align}/
where
$n_{0}(\varepsilon_A, \varepsilon_B)$,
$n_{1}(\varepsilon_A, \varepsilon_B)$,
and
$n_{{\rm Noisy}}(\varepsilon_A, \varepsilon_B)$
are explicitly known
functions obeying
\begin{align}
    &0 \leq
n_{0}(\varepsilon_A, \varepsilon_B),\,
n_{1}(\varepsilon_A, \varepsilon_B),\,
n_{{\rm Noisy}}(\varepsilon_A, \varepsilon_B) \leq 1,\\
&n_{0}(\varepsilon_A, \varepsilon_B)+
n_{1}(\varepsilon_A, \varepsilon_B)
+n_{{\rm Noisy}}(\varepsilon_A, \varepsilon_B)
=1.
\end{align}
Then we also have
\begin{align}
    \va{\overline{W_{{\rm TPM}} (\varepsilon_A, \varepsilon_B)}}
    &\equiv
    n_{0}(\varepsilon_A, \varepsilon_B)
    \va{\overline{W}}_{D}
    +
    n_{1}(\varepsilon_A, \varepsilon_B)
    \va{\overline{W_{{\rm Proj}}}}
    +
    [1-n_{0}(\varepsilon_A, \varepsilon_B)-n_{1}(\varepsilon_A, \varepsilon_B)]
    \va{\overline{W_{{\rm Noisy}}}},
\end{align}
where
\begin{align}
    \va{\overline{W_{{\rm Proj}}}}
    &\equiv \frac{1}{n_{0}(\varepsilon_A, \varepsilon_B)} \Upsilon_{{\rm Proj}},\\
    \va{\overline{W_{{\rm Noisy}}}}
    &\equiv \frac{1}{1-n_{0}(\varepsilon_A, \varepsilon_B)-n_{1}(\varepsilon_A, \varepsilon_B)} \Upsilon_{{\rm Noisy}}.
\end{align}
}\\

\textbf{Remark.}
In the case of symmetric errors, $\varepsilon_A=\varepsilon_B=\varepsilon$, we arrive at Observation 3 in the main text.
\\

\textbf{Remark.}
For any $\varepsilon_A,\varepsilon_B$ and any dimension $d$, we find the inequality
\begin{align}
    \va{\overline{W_{{\rm TPM}} (\varepsilon_A, \varepsilon_B)}}
    \leq
    \va{\overline{W}}_{D},
\end{align}
which is saturated by the limit $\varepsilon_A,\varepsilon_B \to 0.$

{To see this, we first show that
\begin{align}
    p_A^2
    =\sum_i (p_i^A)^2
    =\sum_i \tr[\Pi_i^A \varrho_A]^2
    =\sum_i \tr[\Pi_i^A \varrho_A \Pi_i^A \varrho_A]
    \leq
    \sum_i \tr[\Pi_i^A \varrho_A^2]
    =\tr[\varrho_A^2],
\end{align}
where we employ that
$
    \tr[ABAB] \leq \tr[A^2B^2],
$
for any Hermitian operators $A,B$.
This result directly yields
$dp_A^2-1 \leq r_A^2$.}
Similarly we can have that
$dp_B^2-1 \leq r_B^2$ and
$d^2 p_{AB}^2 - dp_A^2 -dp_B^2 +1 \leq t^2$.
Also, we find
\begin{align} \nonumber
    \sum_{a,b,c}t_{ab} t_{cb} \zeta_{ac}^A
    &=\frac{1}{d^2}\sum_{a,b,c,d}\sum_i t_{ab} t_{cd} 
    \tr(\Pi_{i}^{A} \lambda_a \Pi_{i}^{A} \lambda_c ) \tr(\lambda_b \lambda_d)\\ \nonumber
    &=\frac{1}{d^2}\sum_{a,b,c,d}\sum_i t_{ab} t_{cd} 
    \tr\left[(\Pi_{i}^{A}\otimes \eins_B)
    (\lambda_a \otimes \lambda_b)
    (\Pi_{i}^{A}\otimes \eins_B)
    (\lambda_c \otimes \lambda_d)
    \right]\\ \nonumber
    &=\frac{1}{d^2}\sum_i 
    \tr\left[(\Pi_{i}^{A}\otimes \eins_B) T_2
    (\Pi_{i}^{A}\otimes \eins_B) T_2
    \right]\\
    &\leq \frac{1}{d^2}\sum_i 
    \tr\left[(\Pi_{i}^{A}\otimes \eins_B) T_2^2
    \right]
    = \frac{1}{d^2} \tr[T_2^2] = t^2,
\end{align}
where we employ that
$
    \tr[ABAB] \leq \tr[A^2B^2],
$
for any Hermitian operators $A,B$.
Similarly, we have that $\sum_{a,b,c}t_{ab} t_{ac} \zeta_{bc}^B \leq t^2$.
Substituting these results into the expression
$\va{\overline{W_{{\rm TPM}} (\varepsilon_A, \varepsilon_B)}}$
given in Observation 6 and
using the condition
$f_{\varepsilon_X}^2+2f_{\varepsilon_X}g_{\varepsilon_X}+dg_{\varepsilon_X}^2=1$ 
for $X=A,B$, we can straightforwardly complete the proof.

\subsection{Proof of Observation 5}
We begin by writing the TPM work average for a fixed unitary as
\begin{align} \nonumber
    W_{{\rm TPM}} (\varepsilon_A, \varepsilon_B)
    &=  \sum_{i,j,k,l} m_{ij}m_{kl|ij} w_{ijkl}\\ \nonumber
    &=  \sum_{i,j} m_{ij} e_{ij} - \sum_{i,j,k,l} m_{ij} m_{kl|ij} e_{kl}^\prime\\ \nonumber
    &= \sum_{i,j} \tr[P_i^A \otimes P_j^B \vr_{AB}]e_{ij}
    - \sum_{i,j,k,l} m_{ij} \tr[P_k^A \otimes P_l^B \sigma_{ij}^\prime] e_{kl}^\prime\\
    &= \tr[\vr_{AB} H_{D}] - \sum_{i,j} \tr\left[
    (U_A\otimes U_B)
    \sqrt{P_i^A}\otimes \sqrt{P_j^B} \vr_{AB}
    \sqrt{P_i^A}\otimes \sqrt{P_j^B}
    (U_A\otimes U_B)^\dagger H_{D} \right],
\end{align}
by virtue of Eq.~(\ref{instrumentproperty}).
In order to derive the unitary average $\overline{W_{{\rm TPM}} (\varepsilon_A, \varepsilon_B)}$, we first note that
\begin{align}
    \sqrt{P_i^A} = f_{\varepsilon_A} \Pi_i^A +  g_{\varepsilon_A} \eins_A,\,\,\,\,\,\,\,\,\,
    \sqrt{P_i^B} = f_{\varepsilon_B} \Pi_i^B +  g_{\varepsilon_B} \eins_B,
\end{align}
with $\eins_X = \Pi_i^X + \sum_{j \neq i} \Pi_j^X $. We then define
\begin{align}
    f_{\varepsilon_X} \equiv \sqrt{\varepsilon_X + \frac{1-\varepsilon_X}{d}} -\sqrt{\frac{1-\varepsilon_X}{d}},\,\,\,\,\,\,\,\,\,
    g_{\varepsilon_X} \equiv \sqrt{\frac{1-\varepsilon_X}{d}},
\end{align}
with the normalization condition
\begin{align}\label{normalfge}
    f_{\varepsilon_X}^2+2f_{\varepsilon_X}g_{\varepsilon_X}+dg_{\varepsilon_X}^2=1.
\end{align}
Abbreviating $\Pi_{ij}^{AB} \equiv \Pi_i^A \otimes \Pi_j^B$, a straightforward calculation leads to
\begin{align}
    &\sum_{i,j}
    \sqrt{P_i^A}\otimes \sqrt{P_j^B} \vr_{AB}
    \sqrt{P_i^A}\otimes \sqrt{P_j^B}
    =
    f_{\varepsilon_A}^2 f_{\varepsilon_B}^2 \xi_{AB}
    +\kappa_A \xi_{A}
    +\kappa_B\xi_{B}
    +\kappa_{AB} \vr_{AB},
\end{align}
where we define
\begin{align}\label{xiABxiAxiB}
    \xi_{AB} &\equiv \sum_{i,j} \Pi_{ij}^{AB} \vr_{AB} \Pi_{ij}^{AB},\,\,\,\,\,\,\,\,\,
    \xi_{A} \equiv \sum_{i} \Pi_{i}^{A} \otimes \eins_B \vr_{AB} \Pi_{i}^{A} \otimes \eins_B,\,\,\,\,\,\,\,\,\,
    \xi_{B} \equiv \sum_{j} \eins_A \otimes \Pi_{j}^{B} \vr_{AB} \eins_A \otimes \Pi_{j}^{B},\\
    \kappa_{AB}&\equiv
    {\kappa_A \kappa_B}/({f_{\varepsilon_A}^2 f_{\varepsilon_B}^2}),
    \,\,\,\,\,\,\,\,\,\,\,\,\,
    \kappa_A \equiv
    f_{\varepsilon_A}^2 g_{\varepsilon_B}
    \left(2f_{\varepsilon_B}+dg_{\varepsilon_B}
    \right),\,\,\,\,\,\,\,\,\,\,\,\,\,\,\,\,\,\,\,\,
    \kappa_B \equiv 
    f_{\varepsilon_B}^2 g_{\varepsilon_A}
    \left(2f_{\varepsilon_A}+dg_{\varepsilon_A}
    \right).
\end{align}

From this follows
\begin{align}
    \overline{W_{{\rm TPM}} (\varepsilon_A, \varepsilon_B)}
    &=\tr[\vr_{AB} H_{D}] - \int dU_A \int dU_B  \,
    \tr\left[
    \left(
    f_{\varepsilon_A}^2 f_{\varepsilon_B}^2 \xi_{AB}^\prime
    +\kappa_A \xi_{A}^\prime
    +\kappa_B\xi_{B}^\prime
    +\kappa_{AB} \vr_{AB}^\prime
    \right)
    H_{D} \right],
\end{align}
where
$\chi^\prime = (U_A \otimes U_B) \chi (U_A^\dagger \otimes U_B^\dagger)$
for any $\chi =\xi_{AB}, \xi_{A}, \xi_{B}, \vr_{AB}$.
With help of Eq.~(\ref{Phi1map}) in Appendix \ref{app:A}, we straightforwardly arrive at
\begin{align} \label{workintegral}
    \int dU_A \int dU_B  \, \tr\left[\chi^\prime H_{D} \right]
    = \frac{\tr[H_{D}]}{d^2},
\end{align}
provided that $\tr[\chi^\prime]=1$.
Finally, by applying the normalization condition in Eq.~(\ref{normalfge}), the proof of Observation 5 is completed.

\subsection{Proof of Observation 6}
We begin by recalling that
$
\va{\overline{W_{{\rm TPM}} (\varepsilon_A, \varepsilon_B)}}
    =
    \overline{W_{{\rm TPM}} (\varepsilon_A, \varepsilon_B)^2}
    -\overline{W_{{\rm TPM}} (\varepsilon_A, \varepsilon_B)}^2.
$
Based on the assumption $\tr[H_{D}]=0$ and the result of Observation 5, the second term simplifies to
\begin{align}
    \overline{W_{{\rm TPM}} (\varepsilon_A, \varepsilon_B)}^2
    = \tr \left[\vr_{AB} H_{D}\right]^2.
\end{align}
For the first term $\overline{W_{{\rm TPM}} (\varepsilon_A, \varepsilon_B)^2}$, let us consider the expansion
\begin{align} \nonumber
    \overline{W_{{\rm TPM}} (\varepsilon_A, \varepsilon_B)^2}
    &=\int dU_A \int dU_B  \,
    \left\{
    \tr[\vr_{AB} H_{D}]
    -\tr\left[
    \left(
    f_{\varepsilon_A}^2 f_{\varepsilon_B}^2 \xi_{AB}^\prime
    +\kappa_A \xi_{A}^\prime
    +\kappa_B\xi_{B}^\prime
    +\kappa_{AB} \vr_{AB}^\prime
    \right)
    H_{D} \right]
    \right\}^2\\ \nonumber
    &=\tr[\vr_{AB} H_{D}]^2
    + \int dU_A \int dU_B  \,
    \left\{
    \tr\left[
    \left(
    f_{\varepsilon_A}^2 f_{\varepsilon_B}^2 \xi_{AB}^\prime
    +\kappa_A \xi_{A}^\prime
    +\kappa_B\xi_{B}^\prime
    +\kappa_{AB} \vr_{AB}^\prime
    \right)
    H_{D} \right]
    \right\}^2\\
    &\quad
    -2\tr[\vr_{AB} H_{D}]\int dU_A \int dU_B  \,
    \left\{
    \tr\left[
    \left(
    f_{\varepsilon_A}^2 f_{\varepsilon_B}^2 \xi_{AB}^\prime
    +\kappa_A \xi_{A}^\prime
    +\kappa_B\xi_{B}^\prime
    +\kappa_{AB} \vr_{AB}^\prime
    \right)
    H_{D} \right]
    \right\}.
\end{align}
By virtue of Eq.~(\ref{workintegral}) and the assumption $\tr[H_{D}]=0$, the third line vanishes.
Expanding the second term in the second line, we identify ten types of unitary integrals,
\begin{align}
    \Xi_{\xi_{AB}}
    &= f_{\varepsilon_A}^4 f_{\varepsilon_B}^4
    \int dU_A \int dU_B  \, \tr\left[\xi_{AB}^\prime H_{D} \right]^2,\\
    \Xi_{\xi_{A}}
    &= \kappa_A^2 \int dU_A \int dU_B  \, \tr\left[\xi_{A}^\prime H_{D} \right]^2,\\
    \Xi_{\xi_{B}}
    &= \kappa_B^2 \int dU_A \int dU_B  \, \tr\left[\xi_{B}^\prime H_{D} \right]^2,\\
    \Xi_{\vr_{AB}}
    &= \kappa_{AB}^2 \int dU_A \int dU_B  \, \tr\left[\vr_{AB}^\prime H_{D} \right]^2,\\
    \Xi_{c_1}
    &= f_{\varepsilon_A}^2 f_{\varepsilon_B}^2 \kappa_A
    \int dU_A \int dU_B  \, \tr\left[\xi_{AB}^\prime H_{D} \right] \cdot \tr\left[\xi_{A}^\prime H_{D} \right],\\
    \Xi_{c_2}
    &= f_{\varepsilon_A}^2 f_{\varepsilon_B}^2 \kappa_B
    \int dU_A \int dU_B  \, \tr\left[\xi_{AB}^\prime H_{D} \right] \cdot \tr\left[\xi_{B}^\prime H_{D} \right],\\
    \Xi_{c_3}
    &= \kappa_A \kappa_B
    \int dU_A \int dU_B  \, \tr\left[\xi_{A}^\prime H_{D} \right] \cdot \tr\left[\xi_{B}^\prime H_{D} \right],\\
    \Xi_{c_4}
    &= f_{\varepsilon_A}^2 f_{\varepsilon_B}^2 \kappa_{AB}
    \int dU_A \int dU_B  \, \tr\left[\xi_{AB}^\prime H_{D} \right] \cdot \tr\left[\vr_{AB}^\prime H_{D} \right],\\
    \Xi_{c_5}
    &= \kappa_A \kappa_{AB}
    \int dU_A \int dU_B  \, \tr\left[\xi_{A}^\prime H_{D} \right] \cdot \tr\left[\vr_{AB}^\prime H_{D} \right],\\
    \Xi_{c_6}
    &= \kappa_B \kappa_{AB}
    \int dU_A \int dU_B  \, \tr\left[\xi_{B}^\prime H_{D} \right] \cdot \tr\left[\vr_{AB}^\prime H_{D} \right].
\end{align}
Hence we have
\begin{align} \nonumber
    \va{\overline{W_{{\rm TPM}} (\varepsilon_A, \varepsilon_B)}}
    &=
    \overline{W_{{\rm TPM}} (\varepsilon_A, \varepsilon_B)^2}
    -\overline{W_{{\rm TPM}} (\varepsilon_A, \varepsilon_B)}^2\\
    &= \Xi_{\xi_{AB}}
    + \Xi_{\xi_{A}}
    + \Xi_{\xi_{B}}
    + \Xi_{\vr_{AB}}
    + 2\sum_{i=1}^6 \Xi_{c_i}.
\end{align}
We notice that the fourth term $\Xi_{\vr_{AB}}/\kappa_{AB}^2$ is equal to the theoretical work variance $\va{\overline{W}}_{D}$ for the diagonal Hamiltonian $H_D$ in Observation 1 in the main text.
The first three terms, $\Xi_{\xi_{AB}}, \Xi_{\xi_{A}}, \Xi_{\xi_{B}}$,
can be attributed to the noiseless local TPM, and their sum corresponds to the variance $\va{\overline{W_{{\rm Proj}}}}$ in Observation 3 in the main text.
Finally, all the cross terms $\Xi_{c_i}$ for $i=1,6$ vanish in the limits $\varepsilon_A,\varepsilon_B \to 0,1$; they constitute the additional noise contribution $\va{\overline{W_{{\rm Noisy}}}}$ in Observation 3.

In order to find the explicit form of $\va{\overline{W_{{\rm TPM}} (\varepsilon_A, \varepsilon_B)}}$, we must evaluate all these terms.
We begin by recalling the generalized Bloch representation of $\varrho_{AB}$:
\begin{align}
    \vr_{AB} &=\frac{1}{d^2}\left(
    \eins_{AB}
    + R_1^A \otimes \eins_B
    + \eins_A \otimes R_1^B
    + T_2
    \right),
\end{align}
introducing the traceless Hermitian operators
\begin{align}
    R_1^A&=\sum_{i=1}^{d^2-1} r_i^{A} \lambda_i,\,\,\,\,\,\,~
    R_1^B=\sum_{i=1}^{d^2-1} r_i^{B} \lambda_i,\,\,\,\,\,\,~
    T_2=\sum_{i,j=1}^{d^2-1} t_{ij} \lambda_i \otimes \lambda_j.
\end{align}
For these expressions, we define the quantities
\begin{align}
    r_A^2 &= \frac{1}{d}\tr\left[(R_1^A)^2 \right] = \sum_{i=1}^{d^2-1} (r_i^{A})^2,\,\,\,\,\,\,~
    r_B^2 = \frac{1}{d}\tr\left[(R_1^B)^2 \right] = \sum_{i=1}^{d^2-1} (r_i^{B})^2,\,\,\,\,\,\,~
    t^2   = \frac{1}{d^2}\tr\left[T_2^2 \right] = \sum_{i,j=1}^{d^2-1} t_{ij}^2,
\end{align}
which capture the magnitude of the one- and two-body quantum correlations of $\varrho_{AB}$.
With these expressions, we rewrite the state $\xi_{AB}$ in Eq.~(\ref{xiABxiAxiB}) as
\begin{align}
\xi_{AB} &= \sum_{i,j} \Pi_{ij}^{AB} \vr_{AB} \Pi_{ij}^{AB}
          =\sum_{i,j} p_{ij}^{AB} \Pi_{ij}^{AB},
\end{align}
where
$
    p_{ij}^{AB}
    \equiv \tr[\Pi_{ij}^{AB} \vr_{AB}]
    = ({1}/{d^2})\left\{1+\tr[\Pi_{i}^{A} R_1^A]+\tr[\Pi_{j}^{B}R_1^B]+\tr[\Pi_{ij}^{AB}T_2]\right\}.
$
Also, we define 
$
    p_{i}^A
    \equiv \sum_{j} p_{ij}^{AB}
$
and
$
    p_{j}^B
    \equiv \sum_{i} p_{ij}^{AB}.
$

Letting 
$p_A^2 =\sum_{i} (p_{i}^A)^2$,
$p_B^2 =\sum_{j} (p_{j}^B)^2$,
$p_{AB}^2 = \sum_{i,j} (p_{ij}^{AB})^2$, and
employing the formulas (\ref{Phi1map}, \ref{Phi2map}),
a long calculation leaves us with
\begin{align}
    \Xi_{\xi_{AB}}
    &=  \frac{f_{\varepsilon_A}^4 f_{\varepsilon_B}^4}{d^2-1}
    \left[
    \left(dp_A^2-1\right)h_A^2
    +\left(dp_B^2-1\right)h_B^2
    +\frac{\left(d^2p_{AB}^2 -dp_A^2 - dp_B^2 +1 \right)g^2v^2}{d^2-1}
    \right],\\
    \Xi_{\xi_{A}}
    &=\frac{\kappa_A^2}{d^2-1}
    \left[\left(dp_A^2-1\right)h_A^2 + r_B^2h_B^2
    +\frac{g^2v^2}{d^2-1}\sum_{a,b,c}t_{ab}t_{cb}\zeta_{ac}^A
    \right],
    \\
    \Xi_{\xi_{B}}
    &=\frac{\kappa_B^2}{d^2-1}
    \left[r_A^2h_A^2 + \left(dp_B^2-1\right)h_B^2
    +\frac{g^2v^2}{d^2-1}\sum_{a,b,c}t_{ab}t_{ac}\zeta_{bc}^B
    \right],\\
    \Xi_{c_1}
    &=\frac{f_{\varepsilon_A}^2 f_{\varepsilon_B}^2 \kappa_A}{d^2-1}
    \left[\left(dp_A^2-1\right)h_A^2
    +\left(dp_B^2-1\right)h_B^2
    +\frac{\left(d^2p_{AB}^2 -dp_A^2 - dp_B^2 +1 \right)g^2v^2}{d^2-1}
    \right],
    \\
    \Xi_{c_2}
    &=\frac{f_{\varepsilon_A}^2 f_{\varepsilon_B}^2 \kappa_B}{d^2-1}
    \left[\left(dp_A^2-1\right)h_A^2
    +\left(dp_B^2-1\right)h_B^2
    +\frac{\left(d^2p_{AB}^2 -dp_A^2 - dp_B^2 +1 \right)g^2v^2}{d^2-1}
    \right],
    \\
    \Xi_{c_3}
    &=
    \frac{\kappa_A \kappa_B}{d^2-1}
    \left[\left(dp_A^2-1\right)h_A^2
    +\left(dp_B^2-1\right)h_B^2
    +\frac{(d^2p_{AB}^2 - d{p_A^2} - d{p_B^2} + 1)
    g^2v^2}{d^2-1}
    \right],\\
    \Xi_{c_4}
    &=\frac{f_{\varepsilon_A}^2 f_{\varepsilon_B}^2 \kappa_{AB}}{d^2-1}
    \left[\left(dp_A^2-1\right)h_A^2
    +\left(dp_B^2-1\right)h_B^2
    +\frac{\left(d^2p_{AB}^2 -dp_A^2 - dp_B^2 +1 \right)g^2v^2}{d^2-1}
    \right],
    \\
    \Xi_{c_5}
    &=\frac{\kappa_A \kappa_{AB}}{d^2-1}
    \left[\left(dp_A^2-1\right)h_A^2 + r_B^2h_B^2
    +\frac{g^2v^2}{d^2-1}\sum_{a,b,c}t_{ab}t_{cb}\zeta_{ac}^A
    \right],
    \\
    \Xi_{c_6}
    &=\frac{\kappa_B \kappa_{AB}}{d^2-1}
    \left[r_A^2h_A^2 + \left(dp_B^2-1\right)h_B^2
    +\frac{g^2v^2}{d^2-1}\sum_{a,b,c}t_{ab}t_{ac}\zeta_{bc}^B
    \right].
\end{align}
Here we introduced
\begin{align}
        \zeta_{ab}^A
        \equiv
        \sum_i \frac{\tr(\Pi_{i}^{A} \lambda_a \Pi_{i}^{A} \lambda_b )}{d},\,\,\,\,\,\,\,\,\,
        \zeta_{ab}^B
        \equiv
        \sum_i \frac{\tr(\Pi_{i}^{B} \lambda_a \Pi_{i}^{B} \lambda_b )}{d}.
\end{align}
Summarizing these terms, we can complete the proof of Observation 6.
Here, it might be useful for some readers to note that
\begin{align} \nonumber
    \sum_{a,b,c,d} t_{ab}t_{cd} \zeta_{ac}^A \zeta_{bd}^B
    &=\frac{1}{d^2}\sum_{i,j}\tr[\Pi_{ij}^{AB}T_2\Pi_{ij}^{AB}T_2]\\
    \nonumber
    &=\frac{1}{d^2}\sum_{i,j}\{\tr[\Pi_{ij}^{AB}T_2]\}^2\\
    \nonumber
    &=\frac{1}{d^2}\sum_{i,j}\left(d^2p_{ij}^{AB} -d{p_{i}^A}  -d{p_{j}^B} + 1\right)^2\\
    &=d^2p_{AB}^2 - d{p_A^2} - d{p_B^2} + 1,
\end{align}
where we use that
$\tr[\Pi_{ij}^{AB}T_2] = d^2p_{ij}^{AB} -d{p_{i}^A}  -d{p_{j}^B} + 1$.


\begin{thebibliography}{106}


\bibitem{goold2016role}
J. Goold, M. Huber, A. Riera, L. del Rio, and P. Skrzypczyk,
J. Phys. A {\bf 49}, 143001 (2016).

\bibitem{mukherjee2016presence}
A. Mukherjee, A. Roy, S. S. Bhattacharya, and M. Banik,
Phys. Rev. E {\bf 93}, 052140 (2016).

\bibitem{alimuddin2019bound}
M. Alimuddin, T. Guha, and P. Parashar,
Phys. Rev. A {\bf 99}, 052320 (2019).

\bibitem{bruschi2015thermodynamics}
D. E. Bruschi, M. Perarnau-Llobet, N. Friis, K. V. Hovhannisyan, and M. Huber,
Phys. Rev. E {\bf 91}, 032118 (2015).

\bibitem{perarnau2015extractable}
M. Perarnau-Llobet, K. V. Hovhannisyan, M. Huber, P. Skrzypczyk, N. Brunner, and A. Ac{\'\i}n,
Phys. Rev. X {\bf 5}, 041011 (2015).

\bibitem{huber2015thermodynamic}
M. Huber, M. Perarnau-Llobet, K. V. Hovhannisyan, P. Skrzypczyk, C. Kl{\"o}ckl, N. Brunner, and A. Ac{\'\i}n,
New J. Phys. {\bf 17}, 065008 (2015).

\bibitem{friis2016energetics}
N. Friis, M. Huber, and M. Perarnau-Llobet,
Phys. Rev. E {\bf 93}, 042135 (2016).

\bibitem{brunelli2017detecting}
M. Brunelli, M. G. Genoni, M. Barbieri, and M. Paternostro,
Phys. Rev. A {\bf 96}, 062311 (2017).

\bibitem{mckay2018fluctuations}
E. McKay, N. A. Rodr{\'\i}guez-Briones, and E. Mart{\'\i}n-Mart{\'\i}nez,
Phys. Rev. E {\bf 98}, 032132 (2018).

\bibitem{brunner2014entanglement}
N. Brunner, M. Huber, N. Linden, S. Popescu, R. Silva, and P. Skrzypczyk,
Phys. Rev. E {\bf 89}, 032115 (2014).

\bibitem{brask2015autonomous}
J. B. Brask, G. Haack, N. Brunner, and M. Huber,
New J. Phys. {\bf 17}, 113029 (2015).

\bibitem{campaioli2018quantum}
F. Campaioli, F. A. Pollock, and S. Vinjanampathy,
{\it Thermodynamics in the Quantum Regime, Fundamental Aspects and New Directions}
(Springer, Cham, Switzerland, 2018).

\bibitem{alicki2013entanglement}
R. Alicki and M. Fannes,
Phys. Rev. E {\bf 87}, 042123 (2013).

\bibitem{hovhannisyan2013entanglement}
K. V. Hovhannisyan, M. Perarnau-Llobet, M. Huber, and A. Ac{\'\i}n,
Phys. Rev. Lett. {\bf 111}, 240401 (2013).

\bibitem{salvia2022optimal}
R. Salvia, G. De Palma, and V. Giovannetti,
Phys. Rev. A {\bf 107}, 012405 (2023).

\bibitem{binder2015quantacell}
F. C. Binder, S. Vinjanampathy, K. Modi, and J. Goold,
New J. Phys. {\bf 17}, 075015 (2015).

\bibitem{campaioli2017enhancing}
F. Campaioli, F. A. Pollock, F. C. Binder, L. C{\'e}leri, J. Goold, S. Vinjanampathy, and K. Modi,
Phys. Rev. Lett. {\bf 118}, 150601 (2017).

\bibitem{friis2018precision}
N. Friis and M. Huber,
Quantum {\bf 2}, 61 (2018).

\bibitem{le2018spin}
T. P. Le, J. Levinsen, K. Modi, M. M. Parish, and F. A. Pollock,
Phys. Rev. A {\bf 97}, 022106 (2018).

\bibitem{Ferraro2018highpower}
D. Ferraro, M. Campisi, G. M. Andolina, V. Pellegrini, and M. Polini,
Phys. Rev. Lett. {\bf 120}, 117702 (2018).

\bibitem{Andolina2019extractablework}
G. M. Andolina, M. Keck, A. Mari, M. Campisi, V. Giovannetti, and M. Polini,
Phys. Rev. Lett. {\bf 122}, 047702 (2019).

\bibitem{julia2020bounds}
S. Juli{\`a}-Farr{\'e}, T. Salamon, A. Riera, M. N. Bera, and M. Lewenstein,
Phys. Rev. Research {\bf 2}, 023113 (2020).

\bibitem{Quach2020usingdark}
J. Q. Quach and W. J. Munro,
Phys. Rev. Applied {\bf 14}, 024092 (2020).

\bibitem{Gyhm2022quantumcharging}
J.-Y. Gyhm, D. {\v{S}}afr{\'a}nek, and D. Rosa,
Phys. Rev. Lett. {\bf 128}, 140501 (2022).

\bibitem{quach2022superabsorption}
J. Q. Quach, K. E. McGhee, L. Ganzer, D. M. Rouse, B. W.
Lovett, E. M. Gauger, J. Keeling, G. Cerullo, D. G. Lidzey, and
T. Virgili,
Sci. Adv. {\bf 8}, eabk3160 (2022).

\bibitem{Hu2021Optimal}
C.-K. Hu, J. Qiu, P. J. P. Souza, J. Yuan, Y. Zhou, L. Zhang, J. Chu, X. Pan, L. Hu, J. Li, Y. Xu, Y. Zhong, S. Liu, F. Yan, D. Tan, R. Bachelard, C. J. Villas-Boas, A. C. Santos, and D. Yu,
Quantum Sci. Tech. {\bf 7}, 045018 (2022).

\bibitem{van2012measuring}
S. J. van Enk and C. W. J. Beenakker,
Phys. Rev. Lett. {\bf 108}, 110503 (2012).

\bibitem{tran2015quantum}
M. C. Tran, B. Daki{\'c}, F. Arnault, W. Laskowski, and T. Paterek,
Phys. Rev. A {\bf 92}, 050301(R) (2015).

\bibitem{tran2016correlations}
M. C. Tran, B. Daki{\'c}, W. Laskowski, and T. Paterek,
Phys. Rev. A {\bf 94}, 042302 (2016).

\bibitem{elben2018renyi}
A. Elben, B. Vermersch, M. Dalmonte, J. I. Cirac, and P. Zoller,
Phys. Rev. Lett. {\bf 120}, 050406 (2018).

\bibitem{elben2019statistical}
A. Elben, B. Vermersch, C. F. Roos, and P. Zoller,
Phys. Rev. A {\bf 99}, 052323 (2019).

\bibitem{elben2020cross}
A. Elben, B. Vermersch, R. van Bijnen, C. Kokail, T. Brydges, C. Maier, M. K. Joshi, R. Blatt, C. F. Roos, and P. Zoller,
Phys. Rev. Lett. {\bf 124}, 010504 (2020).

\bibitem{brydges2019probing}
T. Brydges, A. Elben, P. Jurcevic, B. Vermersch, C. Maier, B. P. Lanyon, P. Zoller, R. Blatt, and C. F. Roos,
Science {\bf 364}, 260 (2019).

\bibitem{ketterer2019characterizing}
A. Ketterer, N. Wyderka, and O. G{\"u}hne,
Phys. Rev. Lett. {\bf 122}, 120505 (2019).

\bibitem{ketterer2020entanglement}
A. Ketterer, N. Wyderka, and O. G{\"u}hne,
Quantum {\bf 4}, 325 (2020).

\bibitem{imai2021bound}
S. Imai, N. Wyderka, A. Ketterer, and O. G{\"u}hne,
Phys. Rev. Lett. {\bf 126}, 150501 (2021).

\bibitem{ketterer2020certifying}
A. Ketterer, S. Imai, N. Wyderka, and O. G{\"u}hne,
Phys. Rev. A {\bf 106}, L010402 (2022).

\bibitem{garcia2017quantum}
I. Garc{\'\i}a-Mata, A. J. Roncaglia, and D. A. Wisniacki,
Phys. Rev. E {\bf 95}, 050102(R) (2017).

\bibitem{lobejko2017work}
M. {\L}obejko, J. {\L}uczka, and P. Talkner
Phys. Rev. E {\bf 95}, 052137 (2017).

\bibitem{chenu2018quantum}
A. Chenu, J. Molina-Vilaplana, and A. del Campo,
Sci. Rep. {\bf 8}, 12634 (2018).

\bibitem{chenu2019work}
A. Chenu, J. Molina-Vilaplana, and A. del Campo,
Quantum {\bf 3}, 127 (2019).

\bibitem{salvia2021distribution}
R. Salvia and V. Giovannetti,
Quantum {\bf 5}, 514 (2021).

\bibitem{caravelli2020random}
F. Caravelli, G. Coulter-De Wit, L. P. Garc{\'\i}a-Pintos, and A. Hamma,
Phys. Rev. Research {\bf 2}, 023095 (2020).

\bibitem{oliviero2021random}
S. F. E. Oliviero, L. Leone, F. Caravelli, and Alioscia Hamma,
SciPost Phys. {\bf 10}, 076 (2021).

\bibitem{gennaro2009relaxation}
G. Gennaro, G. Benenti, and G. M. Palma,
Phys. Rev. A {\bf 79}, 022105 (2009).

\bibitem{de2020quantum}
G. De Chiara and M. Antezza,
Phys. Rev. Research {\bf 2}, 033315 (2020).

\bibitem{Shaghaghi2021extracting}
V. Shaghaghi, G. M. Palma, and G. Benenti,
Phys. Rev. E {\bf 105}, 034101 (2022).

\bibitem{rossini2020quantum}
D. Rossini, G. M. Andolina, D. Rosa, M. Carrega, and M. Polini,
Phys. Rev. Lett. {\bf 125}, 236402 (2020).

\bibitem{rosa2020ultra}
D. Rosa, D. Rossini, G. M. Andolina, M. Polini, and M. Carrega,
J. High Energy Phys. {\bf 2020}, 67 (2020).

\bibitem{jia2020spectral}
Y. Jia and J. J. M. Verbaarschot,
J. High Energy Phys. {\bf 2020}, 193 (2020).



\bibitem{allahverdyan2004maximal}
A. E. Allahverdyan, R. Balian, and T. M. Nieuwenhuizen,
Europhys. Lett. (EPL) {\bf 67}, 565 (2004).

\bibitem{Francica2017daemonic}
G. Francica, J. Goold, F. Plastina, and M. Paternostro,
npj Quantum Inf. {\bf 3}, 12 (2017).

\bibitem{Bernards2019daemonic}
F. Bernards, M. Kleinmann, O. Gühne, and M. Paternostro,
Entropy {\bf 21}, 771 (2019).

\bibitem{Francica2022quantum}
G. Francica
Phys. Rev. E {\bf 105}, L052101 (2022).


\bibitem{goodman2000representations}
R. Goodman and N. R. Wallach,
{\it Representations and Invariants of the Classical Groups}
(Cambridge University Press, Cambridge, England, 1998).

\bibitem{roberts2017chaos}
D. A. Roberts and B. Yoshida,
J. High Energy Phys. {\bf 2017}, 121 (2017).

\bibitem{kliesch2021theory}
M. Kliesch and I. Roth,
PRX Quantum {\bf 2}, 010201 (2021).

\bibitem{kimura2003bloch}
G. Kimura,
Phys. Lett. A {\bf 314}, 339 (2003).

\bibitem{bertlmann2008bloch}
R. A. Bertlmann and P. Krammer,
J. Phys. A: Math. Theor. {\bf 41}, 235303 (2008).

\bibitem{Siewert2022orthogonal}
J. Siewert,
J. Phys. Commun. {\bf 6}, 055014 (2022).

\bibitem{aschauer2003local}
H. Aschauer, J. Calsamiglia, M. Hein, and H. J. Briegel,
Quant. Inf. Comp. {\bf 4}, 383 (2004).

\bibitem{de2011multipartite}
J. I. de Vicente and M. Huber,
Phys. Rev. A {\bf 84}, 062306 (2011).

\bibitem{klockl2015characterizing}
C. Kl\"{o}ckl and M. Huber,
Phys. Rev. A {\bf 91}, 042339 (2015).

\bibitem{wyderka2020characterizing}
N. Wyderka and O. G\"{u}hne,
J. Phys. A: Math. Theor. {\bf 53}, 345302 (2020).

\bibitem{eltschka2020maximum}
C. Eltschka and J. Siewert,
Quantum {\bf 4}, 229 (2020).

\bibitem{nielsen2002quantum}
M. A. Nielsen and I. L. Chuang,
{\it Quantum Computation and Quantum Information}
(Cambridge University Press, Cambridge UK, 2010).

\bibitem{cerf2002security}
N. J. Cerf, M. Bourennane, A. Karlsson, and N. Gisin,
Phys. Rev. Lett. {\bf 88}, 127902 (2002).

\bibitem{barrett2006maximally}
J. Barrett, A. Kent, and S. Pironio,
Phys. Rev. Lett. {\bf 97}, 170409 (2006).

\bibitem{buscemi2011entanglement}
F. Buscemi and N. Datta,
Phys. Rev. Lett. {\bf 106}, 130503 (2011).

\bibitem{huber2013weak}
M. Huber and M. Paw{\l}owski,
Phys. Rev. A {\bf 88}, 032309 (2013).

\bibitem{terhal2000schmidt}
B. M. Terhal and P. Horodecki,
Phys. Rev. A {\bf 61}, 040301(R) (2000).

\bibitem{guhne2009entanglement}
O. G{\"u}hne and G. T{\'o}th,
Phys. Rep. {\bf 474}, 1 (2009).

\bibitem{friis2019entanglement}
N. Friis, G. Vitagliano, M. Malik, and M. Huber,
Nat. Rev. Phys. {\bf 1}, 72 (2019).

\bibitem{sanpera2001schmidt}
A. Sanpera, D. Bru{\ss}, and M. Lewenstein,
Phys. Rev. A {\bf 63}, 050301(R) (2001).

\bibitem{sperling2011determination}
J. Sperling and W. Vogel,
Phys. Rev. A {\bf 83}, 042315 (2011).

\bibitem{sperling2011schmidt}
J. Sperling and W. Vogel,
Phys. Scr. {\bf 83}, 045002 (2011).

\bibitem{huber2018high}
M. Huber, L. Lami, C. Lancien, and A. M{\"u}ller-Hermes,
Phys. Rev. Lett. {\bf 121}, 200503 (2018).

\bibitem{Liu2022bounding}
S. Liu, N. Fadel, Q. He, M. Huber, G. Vitagliano,
{\it Bounding entanglement dimensionality from the covariance matrix},
arXiv:2208.04909

\bibitem{horodecki1996information}
R. Horodecki and M. Horodecki,
Phys. Rev. A {\bf 54}, 1838 (1996).

\bibitem{talkner2007fluctuation}
P. Talkner, E. Lutz, and P. H{\"a}nggi,
Phys. Rev. E {\bf 75}, 050102(R) (2007).

\bibitem{esposito2009nonequilibrium}
M. Esposito, U. Harbola, and S. Mukamel,
Rev. Mod. Phys. {\bf 81}, 1665 (2009).

\bibitem{campisi2011colloquium}
M. Campisi, P. H{\"a}nggi, and P. Talkner,
Rev. Mod. Phys. {\bf 83}, 771 (2011).

\bibitem{roncaglia2014work}
A. J. Roncaglia, F. Cerisola, and J. P. Paz,
Phys. Rev. Lett. {\bf 113}, 250601 (2014).

\bibitem{de2015measuring}
G. De Chiara1, A. J Roncaglia, and J. P. Paz,
New J. Phys. {\bf 17}, 035004 (2015).

\bibitem{talkner2016aspects}
P. Talkner and P. H{\"a}nggi,
Phys. Rev. E {\bf 93}, 022131 (2016).

\bibitem{perarnau2017no}
M. Perarnau-Llobet, E. B{\"a}umer, K. V. Hovhannisyan, M. Huber, and A. Ac{\'\i}n,
Phys. Rev. Lett. {\bf 118}, 070601 (2017).

\bibitem{baumer2018fluctuating}
E. B{\"a}umer, M. Lostaglio, M. Perarnau-Llobet, and R. Sampaio,
Fluctuating work in coherent quantum systems: Proposals and limitations, in
{\it Thermodynamics in the Quantum Regime, Fundamental Aspects and New Directions}
(Springer, Cham, Switzerland, 2018).

\bibitem{de2018ancilla}
G. De Chiara, P. Solinas, F. Cerisola, and A. J. Roncaglia,
Ancilla-Assisted Measurement of Quantum Work, in
{\it Thermodynamics in the Quantum Regime, Fundamental Aspects and New Directions}
(Springer, Cham, Switzerland, 2018).

\bibitem{lostaglio2018quantum}
M. Lostaglio,
Phys. Rev. Lett. {\bf 120}, 040602 (2018).

\bibitem{bednorz2010quasiprobabilistic}
A. Bednorz and W. Belzig,
Phys. Rev. Lett. {\bf 105}, 106803 (2010).

\bibitem{guryanova2020ideal}
Y. Guryanova, N. Friis, and M. Huber,
Quantum {\bf 4}, 222 (2020).

\bibitem{debarba2019work}
T. Debarba, G. Manzano, Y. Guryanova, M. Huber, and N. Friis,
New J. Phys. {\bf 21}, 113002 (2019).

\bibitem{beyer2021joint}
K. Beyer, R. Uola, K. Luoma, and W. T. Strunz,
Phys. Rev. E {\bf 106}, L022101 (2022).

\bibitem{horodecki2002method}
P. Horodecki and A. Ekert,
Phys. Rev. Lett. {\bf 89}, 127902 (2002).

\bibitem{horodecki2003measuring}
P. Horodecki,
Phys. Rev. Lett. {\bf 90}, 167901 (2003).

\bibitem{ekert2002direct}
A. K. Ekert, C. M. Alves, D. K. L. Oi, M. Horodecki, P. Horodecki, and L. C. Kwek,
Phys. Rev. Lett. {\bf 88}, 217901 (2002).

\bibitem{carteret2005noiseless}
H. A. Carteret,
Phys. Rev. Lett. {\bf 94}, 040502 (2005)

\bibitem{bovino2005direct}
F. A. Bovino, G. Castagnoli, A. Ekert, P. Horodecki, C. M. Alves, and A. V. Sergienko,
Phys. Rev. Lett. {\bf 95}, 240407 (2005).

\bibitem{schmid2008experimental}
C. Schmid, N. Kiesel, W. Wieczorek, H. Weinfurter, F. Mintert, and A. Buchleitner,
Phys. Rev. Lett. {\bf 101}, 260505 (2008).

\bibitem{mintert2005concurrence}
F. Mintert, M. Ku{\'s}, and A. Buchleitner,
Phys. Rev. Lett. {\bf 95}, 260502 (2005).

\bibitem{aolita2006measuring}
L. Aolita and F. Mintert,
Phys. Rev. Lett. {\bf 97}, 050501 (2006).

\bibitem{mintert2007observable}
F. Mintert and A. Buchleitner,
Phys. Rev. Lett. {\bf 98}, 140505 (2007).

\bibitem{walborn2007experimental}
S. P. Walborn, P. H. Souto Ribeiro, L. Davidovich, F. Mintert, and A. Buchleitner,
Phys. Rev. A {\bf 75}, 032338 (2007).

\bibitem{kafri2012holevo}
D. Kafri and S. Deffner,
Phys. Rev. A {\bf 86}, 044302 (2012).

\bibitem{gardas2018quantum}
B. Gardas and S. Deffner,
Sci. Rep. {\bf 8}, 17191 (2018).


\bibitem{heinosaari2011mathematical}
T. Heinosaari and M. Ziman,
{\it The mathematical language of quantum theory: from uncertainty to entanglement}
(Cambridge University Press, Cambridge, 2012).
\end{thebibliography}
\end{document}